\documentclass[conference]{IEEEtran}

\usepackage{comment}
\usepackage{dsfont}
\usepackage{cite}
\usepackage{amsmath,amssymb,amsfonts}
\usepackage{algorithmic}
\usepackage{graphicx}
\usepackage{textcomp}

\usepackage{cite}
\usepackage{amsmath,amssymb,amsfonts}
\usepackage{algorithmic}
\usepackage{graphicx}
\usepackage{epstopdf}
\graphicspath{{\figures}}
\usepackage{textcomp}
\usepackage{xcolor}
\usepackage{verbatim}
\usepackage{makecell}
\usepackage{booktabs}
\usepackage{CJK}
\usepackage{multirow}
\usepackage{subcaption} 
\usepackage{caption}
\usepackage{float}      

\usepackage{algorithm}
\usepackage{algorithmic}

\makeatletter
\newcommand{\removelatexerror}{\let\@latex@error\@gobble}
\makeatother
\IEEEoverridecommandlockouts 
\begin{document}
	\begin{CJK}{UTF8}{gbsn}
		\title{\huge Transformer based Collaborative Reinforcement Learning for Fluid Antenna System (FAS)-enabled 3D UAV Positioning}
        \author{Xiaoren Xu, Hao Xu, \emph{Senior Member, IEEE}, Dongyu Wei, Walid Saad, \emph{Fellow, IEEE}, \\ Mehdi Bennis, \emph{Fellow, IEEE}, and Mingzhe Chen, \emph{Senior Member, IEEE}\\
        
			\vspace{-0.6cm}
			
            \thanks{X. Xu and D. Wei are with the Department of Electrical and Computer Engineering, University of Miami, Coral Gables, FL, 33146, USA (Email: xiaoren.xu@miami.edu; dongyu.wei@miami.edu).}
            \thanks{H. Xu is with the Department of Electronic and Electrical Engineering, University College London, London, United Kingdom (Email: hao.xu@ucl.ac.uk).}
            \thanks{W. Saad is with the Wireless@VT Group, Bradley Department of Electrical and Computer Engineering, Virginia Tech, Arlington, VA 24061 USA (Email: walids@vt.edu).}
            \thanks{M. Bennis is with the Centre for Wireless Communications, University of Oulu, 90014 Oulu, Finland (Email: mehdi.bennis@oulu.fi).}
            \thanks{M. Chen is with the Department of Electrical and Computer Engineering, and also with the Frost Institute for Data Science and Computing, University of Miami, Coral Gables, FL, 33146, USA (Email: mingzhe.chen@miami.edu).}

		}
		\maketitle
		\begin{abstract}
			In this paper, a novel Three dimensional (3D) positioning framework of fluid antenna system (FAS)-enabled unmanned aerial vehicles (UAVs) is developed. In the proposed framework, a set of controlled UAVs including an active UAV and four FAS-enabled passive UAVs cooperatively estimate the real-time 3D position of a target UAV. Here, the active UAV transmits a measurement signal to the passive UAVs via the reflection from the target UAV. Each passive UAV estimates the distance of the active-target-passive UAV link and selects an antenna port to share the distance information with the base station (BS), which calculates the real-time position of the target UAV. As the target UAV is moving due to its task operation, the controlled UAVs must optimize their trajectories and select optimal antenna port for transmitting the positioning information, aiming to estimate the real-time position of the target UAV. We formulate this problem as an optimization problem whose goal is to minimize the target UAV positioning error via optimizing the trajectories of all controlled UAVs and antenna port selection of passive UAVs. To address this problem, an attention-based recurrent multi-agent reinforcement learning (AR-MARL) scheme is proposed, which enables each controlled UAV to use the local Q function to determine its trajectory and antenna port while optimizing the target UAV positioning performance without knowing the trajectories and antenna port selections of other controlled UAVs. Different from current MARL methods that use feedforward neural networks to approximate Q functions, the proposed method uses a recurrent neural network (RNN) that incorporates historical state-action pairs of each controlled UAV, and an attention mechanism to analyze the importance of these historical state-action pairs, thus improving the global Q function approximation accuracy and the target UAV positioning accuracy.
            Simulation results show that the proposed AR-MARL scheme can reduce the average positioning error by up to 17.5\% and 58.5\% compared to the VD-MARL scheme and the proposed method without FAS.
			
		\end{abstract}
		

		\section{Introduction}
         Three dimensional (3D) unmanned aerial vechicle (UAV) positioning enables UAVs to function effectively, safely, and autonomously across various tasks (e.g., virtual reality and urgent rescue) \cite{8660516, UAV_Wildfire, RUIZHANG_UAV_POS}. However, realizing accurate 3D UAV positioning faces several challenges. First, different from traditional 2D ground user positioning, UAVs are flying at a high speed in 3D space, making it difficult to estimate their precise locations in real time. Second, signal transmissions over the air are vulnerable to interferences from both UAV and ground communication systems, thus resulting in degraded signal quality and communication disruptions. Third, UAVs often operate in complex and dynamic environments in which obstacles such as buildings can block or reflect signals, causing multipath propagation and further complicating accurate 3D positioning \cite{YH}.

         Recently, several works in \cite{BiSuzhi_UAV,Angela_UAV,10008079, Zhu_UAV,Vehicle-radar,CL_UAVPOS,IOTJ_UAVPOS,RBS_UAVPOS} have studied the optimization of 3D UAV positioning performance. In particular, the authors in \cite{BiSuzhi_UAV} optimized the 3D deployment and resource allocation of UAVs to improve transmission rates while ensuring localization accuracy. The authors in \cite{Angela_UAV} studied the dual-functional UAVs to improve both communication an localization through introducing a new metric that minimized average estimation error of positioning and a low-complexity algorithm. The authors in \cite{10008079} evaluated how different receiving power models affect the performance for positioning fixed-wing UAVs, and the positioning error was reduced significantly by including the polarization factor in the receiving power model. The authors in \cite{Zhu_UAV} used several UAVs to localize a target UAV by optimizing UAV trajectory and transmit power based on the Z function decomposition based reinforcement learning (ZD-RL) method. The authors in \cite{Vehicle-radar} studied a radar assisted vehicle system for precisely tracking the unauthorized UAVs that may cause physical or informational damages. The authors in \cite{CL_UAVPOS} proposed a 3D time difference of arrival (TDOA) positioning scheme to improve the positioning accuracy and computing efficiency for large-scale UAV localization. The authors in \cite{IOTJ_UAVPOS} considered a 2D direction of arrival estimation framework with monostatic multiple input multiple output (MIMO) radar system to localize anonymous UAVs. In \cite{RBS_UAVPOS}, the authors proposed a resonant beam-based approach to improve UAV swarms positioning accuracy while reducing the computational power. However, the current works in \cite{BiSuzhi_UAV,Angela_UAV,10008079, Zhu_UAV,Vehicle-radar,CL_UAVPOS,IOTJ_UAVPOS,RBS_UAVPOS} require all positioning devices to transmit measurement signals to localize the target, and hence, they may not be applied for the scenarios where the UAVs do not have enough power to continuously transmit measurement signals. Meanwhile, the authors in \cite{BiSuzhi_UAV,Angela_UAV,10008079, Zhu_UAV,Vehicle-radar,CL_UAVPOS,IOTJ_UAVPOS,RBS_UAVPOS} did not consider the use of the fluid antenna system (FAS) to improve the performance of 3D UAV positioning. In fact, using FAS, each UAV can select its antenna port according to the observed angle-of-departure (AoD) of multi-path propagation thus significantly improving the positioning accuracy \cite{10599127,NWKCOMST,ISAC_FAS}.

         Recently, several works in \cite{FAS_UAV_Zhang,FAS_UAV_WQQ,FAS_UAV_Cui,FAS_UAV_ISAC,UAV_FAS_WKN,UAV_FAS_RIS,YuBai} have studied the use of fluid antenna (FA)/movable antenna (MA) techniques for UAV-based wireless communications. In particular, the authors in \cite{FAS_UAV_Zhang} employed a six-dimensional (6D) MA array to mitigate the interferences for UAV communications, and jointly optimized the antenna position vector, array rotation vector, receive beamforming vector, and the base station selection through a block coordinate descent (BCD) method. In \cite{FAS_UAV_WQQ}, the authors jointly optimized the UAV deployment, antenna position vector, and antenna weight vector in a MA-aided UAV communication system, where the beamforming gains for secondary users were maximized and the interferences of primary users were minimized. The authors in \cite{FAS_UAV_Cui} studied an MA array-assisted UAV multi-input and single-output communication system for multiple users, in which an alternating optimization (AO) algorithm was developed to maximize the sum data rate of users via adjusting the UAV trajectories, beamforming, and antenna positions. The authors in \cite{FAS_UAV_ISAC} used an MA array-enabled low-altitude platform for integrated sensing and communications applications, where the UAV beamforming and antenna positioning were cooperatively optimized with an AO method to improve the data rate of users while maintaining the required sensing performance. In \cite{UAV_FAS_WKN}, the authors proposed a UAV-relaying FAS with non-orthogonal multiple access (NOMA) technique, in which the active antenna port, UAV height, and power allocation were optimized for the user rate maximization. The authors in \cite{UAV_FAS_RIS} proposed a reconfigurable intelligent surface (RIS) network with FA array-mounted UAVs. To maximize the downlink sum rate, the beamforming of UAV and RIS, UAV deployment, and FA position were optimized using the successive convex approximation (SCA) and sequential rank-one constraint relation (SROCR) methods. The authors in \cite{YuBai} designed a directional MA-assisted UAV backscatter sensor network, where a soft actor critic algorithm was adopted to optimize the trajectory of UAV and direction of MA, thus reducing the data collection time and energy consumption. However, most of these works \cite{FAS_UAV_Zhang,FAS_UAV_WQQ,FAS_UAV_Cui,FAS_UAV_ISAC,UAV_FAS_WKN,UAV_FAS_RIS,YuBai} focused on the optimization of communication performances (e.g., user data rate) instead of UAV sensing and positioning, which requires the collaboration of several sensing devices to simultaneously estimate the distance between the sensing device and the target. Hence, the optimization of FAS for UAV sensing must jointly consider the wireless environment conditions (e.g., wireless channels) and the movement of all sensing devices and the target. Meanwhile, since the target tracked by the sensing UAVs is moving due to its task operation and the sensing UAVs must move accordingly to track the target UAV, the optimization of FAS should also consider the wireless network dynamics introduced by the unknown movement of the sensing UAVs and the target. 
         
         

         The main contribution of our work is a novel 3D UAV positioning framework that enables an active UAV and several FAS-assisted passive UAVs to cooperatively track a target UAV without sharing their observed wireless network information. Our key contributions include:
         \begin{itemize}
         \item We consider a novel FAS-assisted 3D passive UAV positioning framework in which five controlled UAVs including one active UAV and four passive UAVs cooperatively estimate the position of a target UAV. The active UAV transmits a measuring signal to the target UAV, and this signal is received by passive UAVs via the reflection from the target UAV. Each passive UAV estimates the distance of the active-target-passive UAV link. Then, the passive UAV selects an antenna port to share the active-target-passive UAV link distance with the ground base station (BS). Given the distance information, the BS calculates the real-time position of the target UAV.
         \item As the location of the target UAV is always changing due to its task operation, the controlled UAVs must optimize their trajectories and select the optimal antenna port for positioning information transmission, so as to adapt to the dynamic environment and accurately estimate the real-time position of the target UAV. 
         We formulate an optimization problem aiming to minimize the positioning error of the target UAV while satisfying the distance information transmission latency requirements of passive UAVs by optimizing the trajectories and antenna port selections of the controlled UAVs.
         \item To solve the optimization problem, an attention-based recurrent multi-agent reinforcement learning (AR-MARL) scheme is proposed, which enables the controlled UAVs to determine their trajectories and antenna ports through their locally observed information. Different from the value decomposition-based multi-agent reinforcement learning scheme (VD-MARL) \cite{distributedhu}, the proposed method uses a recurrent neural network (RNN) acting as a local Q function of each controlled UAV to capture its historical state-action pairs, and a transformer to analyze the importance of these historical state-action pairs, thus improving the global Q function approximation accuracy, and enhancing the target UAV positioning accuracy.
         \item To understand the optimal target UAV positioning accuracy achieved by the designed framework, we analyze how the positions and the transmit power of UAVs affect the positioning error of the target UAV. The results from our analysis shows that the positioning error can be minimized when the minimum distances between the passive UAVs and the target UAV are equal, and the transmit power of the active UAV is maximized.
         \end{itemize}
         Simulation results show that the proposed AR-MARL scheme can reduce the average positioning error of the target UAV by up to 17.5\%, 31.8\%, and 58.5\% compared to the VD-MARL scheme, independent Q scheme, and the proposed method without FAS. \emph{To the best of our knowledge, this is the first work that uses FAS to optimize the performance of 3D UAV positioning system.}

         The rest of this paper is organized as follows. The proposed FAS-assisted 3D UAV positioning network and problem formulation are described in Section II. The designed AR-MARL scheme for optimizing the trajectories and antenna port selections of the UAVs is introduced in Section III. In Section IV, we analyze the positioning error of the target UAV. Numerical simulation results are presented and analyzed in Section V. Finally, Section VI concludes this paper.

         \begin{table}[t!]
\captionsetup{font=footnotesize} 
\caption{\label{training}LIST OF NOTATIONS}  
\centering 
\setlength{\abovecaptionskip}{0.cm}
\setlength{\tabcolsep}{0.8mm}{
\begin{tabular}{|c|c|}  
\hline  
Notation & Description  \\
\hline  
$\mathcal{K}$ & Set of controlled UAVs \\ 
\hline 
$N$ & Number of predetermined antenna ports \\ 
\hline 
$S$ & Normalized size of the FAS \\ 
\hline 
$\lambda$ & Wavelength \\ 
\hline 
$\boldsymbol{q}_{k,t}$ & 3D position of controlled UAV $k$ at time slot $t$ \\ 
\hline
$\boldsymbol{u}_{t}$ & Actual 3D position of target UAV at time slot $t$ \\ 
\hline
$v$ & Constant moving speed of controlled UAVs \\ 
\hline
$c$ & Speed of light \\ 
\hline
$\psi_{k,t}$ & Yaw angle of controlled UAV $k$ at time slot $t$  \\
\hline
$\theta_{k,t}$ & Pitch angle of controlled UAV $k$ at time slot $t$  \\
\hline
$\tau_{k,t}$ & Transmission time of the measuring signal  \\
\hline
$\omega_{t}$ & Measuring signal from active UAV at time slot $t$  \\
\hline
$\Delta_t$ & Time duration of a time slot \\
\hline
$d_{k,t}$ & Distance from target UAV to each controlled UAV  \\
\hline
$\nu$ & Additive White Guassian Noise \\
\hline
$p_{0,t}$ & Transmission power from active UAV \\
\hline
$p_{k,t}$ & Transmission power from controlled UAV $k$\\
\hline
$\alpha_{0}$ & Path loss in LoS condition at unit distance \\
\hline
$\beta_{k,t}$ & Reflecting coefficient of target UAV \\
\hline
$l_{k,t}$ & Path loss between target UAV and each controlled \\ 
 & UAV at time slot $t$ \\
\hline
$\rho^2$ & Variance of AWGN \\
\hline
$h_{k,t}$ & Received signal by passive UAV $k$ at time slot $t$ \\
\hline
$\gamma_{k,t}^\textrm{P}$ & SNR of the measuring signal for passive UAV $k$ at time slot $t$ \\
\hline
${m}_{k,t}$ & Actual distance of each active-target-passive \\ 
 & UAV link at time slot $t$ \\
\hline
$\hat{m}_{k,t}$ & Estimated distance of each active-target-passive \\ 
 & UAV link at time slot $t$ \\
\hline
$I_{k,t}$ & Number of LoS or NLoS signal
paths between passive UAV $k$ \\ 
& and the ground BS at time slot $t$\\
\hline
$g_{k,t}$ & Channel gain from passive UAV $k$ to ground BS at
time slot $t$ \\
\hline
$\varepsilon_i$ & Fading coefficient of path $i$\\
\hline
$n_{k,t}$ & Active antenna port index of FAS for passive UAV $k$ \\ 
& at time slot $t$\\
\hline
$\varphi_{k,t}^i$ & AoD of path $i$ from passive UAV $k$ at time slot $t$\\
\hline
$L^\textrm{B}$ & Distance-dependent path loss \\
\hline
$L^\textrm{F}$ & Free-space path loss \\
\hline
$r_{k,t}$ & Distance between passive UAV $k$ and BS at time slot $t$ \\
\hline
$\kappa$ & Shadowing random variable \\
\hline
$\eta$ & Variance of shadowing random variable \\
\hline
$\gamma_{k,t}^\textrm{B}$ & SINR of the signal from passive UAV $k$ to BS at time slot $t$ \\
\hline
$W_{k,t}^\textrm{L}$ & Transmission latency of estimated distance information \\
\hline
$D$ &  Data size of the distance information \\
\hline
$B$ &  Bandwidth of the UAV-ground BS transmissions \\
\hline
$e_{k,t}$ & Gaussian error of active-target-passive UAV link \\
& distance measurement at time slot $t$\\
\hline
$\sigma_{k,t}^2$ & Variance of Gaussian measurement error\\
\hline
$\boldsymbol{q}_\textrm{B}$ & 3D position of ground BS \\
\hline
$\hat{\boldsymbol{u}}_t$ & Estimated 3D position of target UAV at time slot $t$ \\
\hline
$T$ & Number of time slots \\
\hline
$L_\textrm{min}$ & Minimum distance restriction between any two UAVs  \\
\hline
$L_\textrm{max}$ & Maximum distance restriction between any two UAVs  \\
\hline
\end{tabular}}    
\vspace{-0.3cm}
\end{table}
		
		\section{System Model and Problem Formulation}
\begin{figure}[t!]
        \setlength\abovecaptionskip{0.cm}
        \setlength\belowcaptionskip{-0.25cm}
        \centering 
        \includegraphics[width=8.5cm]{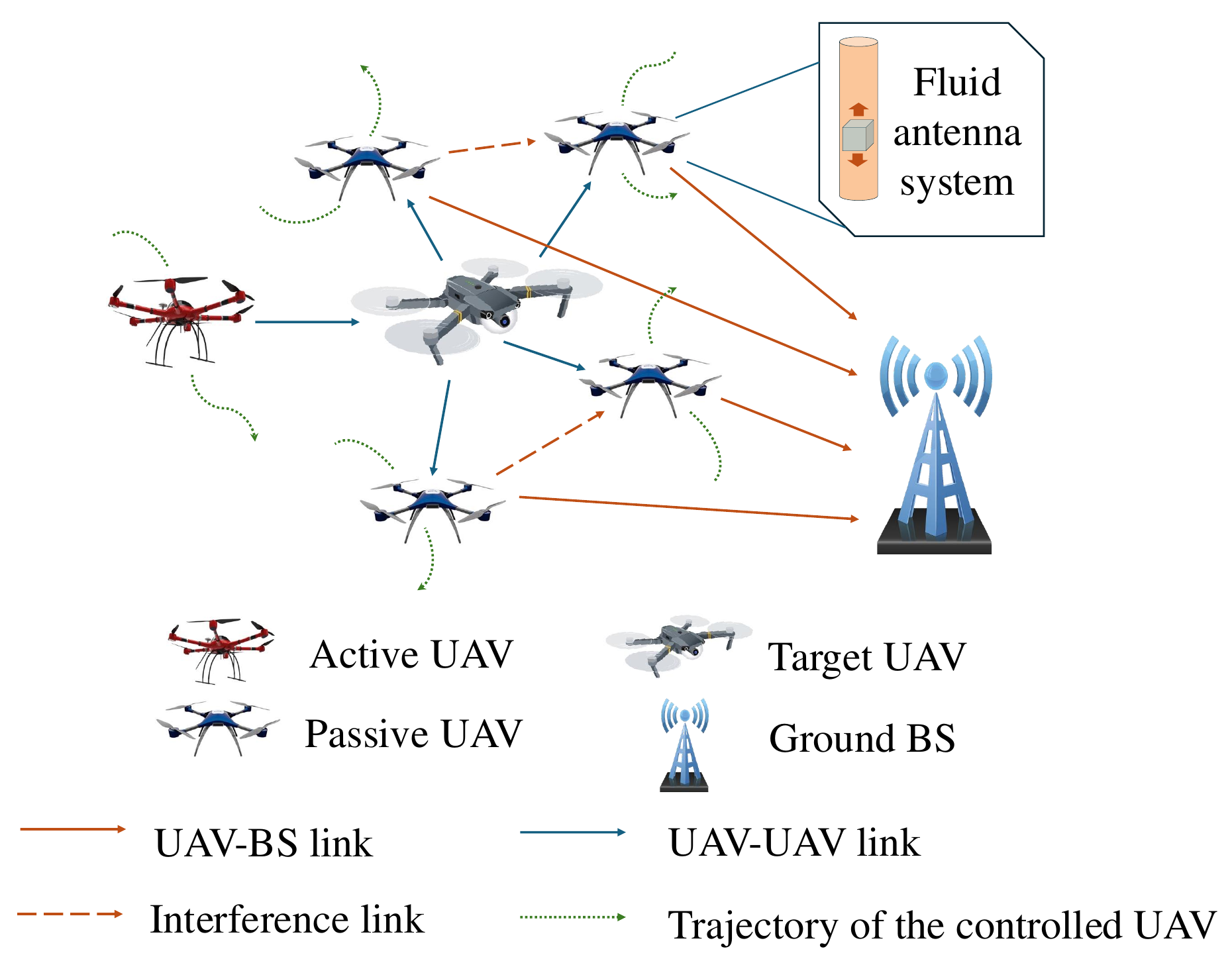}
        \captionsetup{labelsep=period} 
        \caption{FAS-assisted 3D UAV Positioning Network.}
        \label{scenario} 
        \vspace{-0.4cm}
        \end{figure}

		We consider a 3D UAV positioning network in which a set $\mathcal{K}$ of five controlled UAVs including one active UAV and four passive UAVs cooperate with a base station (BS) to estimate the position of a target UAV, as illustrated in Fig. \ref{scenario}. The BS and active UAV use fixed-position antennas.
		In contrast, each passive UAV is equipped with a linear FAS.
		We assume that the location of each FAS can be instantaneously switched to one of the $N$ predetermined antenna ports, which are evenly distributed along a linear dimension of length $S \lambda$ and share a common RF chain\footnote{This structure can be viewed as an approximation of an RF pixel-based linear FAS that has many compact antenna pixels, wherein a single pixel can be activated at a time \cite{new2023information}. This technology enables seamless antenna switching with virtually no time delay.}, with $S$ being the normalized size of the FAS and $\lambda$ being the wavelength. The active UAV transmits a measuring signal to the target UAV, and this signal is received by passive UAVs via the reflection from the target UAV. Here, the passive UAVs will use their fixed antenna ports to receive the signals reflected by the target UAV. Then, each passive UAV will estimate the distance between the active UAV and the target UAV, as well as the distance between the target UAV and itself. Given the estimated distances, the passive UAVs will select another antenna port to transmit the estimated distance information to the BS. Due to the use of FAs, each passive UAV can either change its location or the antenna port to improve the data rate of the UAV-ground BS link in order to improve the localization accuracy and speed. Finally, the BS will use the received distance information to calculate the location of the target UAV. Next, we first introduce the models of UAV movement, data transmissions and positioning. Then, we present the optimization problem.
  
		\subsection{UAV Movement Model}
		Let the 3D location of UAV $k$ at time slot $t$ be $\boldsymbol{q}_{k,t}=\left[x_{k,t},y_{k,t},z_{k,t}\right]^T$, $k\in\mathcal{K}$. Here, $\boldsymbol{q}_{0,t}$ is the location of the active UAV and $\boldsymbol{q}_{k,t}, k={1,2,3,4}$ are the locations of passive UAVs. Then, the real-time 3D location of UAV $k$ is given by
		\begin{equation}\label{UAV 3D Motion}
			\boldsymbol{q}_{k,t+1}\left(\psi_{k,t},\theta_{k,t}\right)=\boldsymbol{q}_{k,t}+v\Delta_t\begin{bmatrix} \cos\psi_{k,t}\cos\theta_{k,t}\\\sin\psi_{k,t}\cos\theta_{k,t}\\\sin\theta_{k,t}
			\end{bmatrix},
		\end{equation}
		where $\psi_{k,t}$ represents the yaw angle, $\theta_{k,t}$ is the pitch angle, $v$ denotes the UAV moving speed, and $\Delta_t$ denotes the time duration of a time slot.
		
		\subsection{Transmission Model} 
		Next, we introduce the models of signal transmission over UAV-UAV links and UAV-ground BS links. 
		\subsubsection{UAV-UAV signal transmission}
		Let $\omega_t \sim {\cal {CN}}(0, 1)$ be the measuring Guassian signal that the active UAV transmits to the target UAV at time slot $t$. We assume that the signal transmission link between the active UAV and the target UAV, as well as the signal transmission link from the target UAV to the passive UAV are Line-of-sight (LoS). The time duration of transmitting measuring signal $\omega_t$ from the active to passive UAV $k$ through the target UAV is
		\begin{equation}
			\tau_{k,t}\left(\boldsymbol{q}_{0,t},\boldsymbol{q}_{k,t}\right)=\frac{d_{0,t}\left(\boldsymbol{q}_{0,t},\boldsymbol{u}_t\right)+d_{k,t}\left(\boldsymbol{u}_{t},\boldsymbol{q}_{k,t}\right)}{c},
		\end{equation}where
		$d_{0,t}\left(\boldsymbol{q}_{0,t},\boldsymbol{u}_t\right)=\Vert \boldsymbol{q}_{0,t}-\boldsymbol{u}_{t} \Vert$ is the distance between the active UAV and the target UAV with $\boldsymbol{u}_t=\left[x_t,y_t,z_t\right]^T$ being the 3D location of the target UAV, $d_{k,t}\left(\boldsymbol{u}_{t},\boldsymbol{q}_{k,t}\right)=\Vert \boldsymbol{u}_{t}-\boldsymbol{q}_{k,t} \Vert$ is the the distance from the target UAV to passive UAV $k$, and $c$ is the speed of light. Since switching the FAS antenna ports cannot increase the channel gain for a UAV-to-UAV LoS link \cite{xu2023capacity}, we assume that each passive UAV $k$ uses a fixed antenna port of its FAS to receive the signal transmitted from the active UAV and reflected by the target UAV, which is expressed as
		\begin{equation}\label{u2u_received signal}
			h_{k,t}\left(\boldsymbol{q}_{0,t},\boldsymbol{q}_{k,t}\right)=\sqrt{p_{0,t}} l_{0,t} l_{k,t} \beta_{k,t} \omega_{t-\tau_{k,t}}+\nu,
		\end{equation} where $p_{0,t}$ is the transmit power of the active UAV, $\beta_{k,t}$ denotes the reflecting coefficient of the target UAV, $l_{0,t}=\sqrt{\alpha_0}d_{0,t}^{-1}\left(\boldsymbol{q}_{0,t},\boldsymbol{u}_{t}\right)$ is the path loss of the active UAV-target UAV link with $\alpha_0$ being the path loss in LoS condition at unit distance (i.e., 1 $m$), $l_{k,t}=\sqrt{\alpha_0}d_{k,t}^{-1}\left(\boldsymbol{q}_{k,t},\boldsymbol{u}_{t}\right)$ is the path loss between the target UAV and passive UAV $k$, and $\nu$ is the Additive White Guassian Noise (AWGN) complying with $\mathcal{CN}(0,\rho^2)$.
		
		Given \eqref{u2u_received signal}, the signal-to-noise ratio (SNR) of the measuring signal for each passive UAV at time slot $t$ is
		\begin{equation}\label{SNR_passiveUAV}
			\gamma^\textrm{P}_{k,t}\left(\boldsymbol{q}_{0,t},\boldsymbol{q}_{k,t}\right)=\frac{p_{0,t}\Vert l_{0,t}l_{k,t}\beta_{k,t}\Vert^2}{\rho^2}.
		\end{equation}

		\subsubsection{UAV-ground BS signal transmission}
		Passive UAVs need to estimate the distance from the active UAV to passive UAV $k$ via the target UAV using the received signals. Then, the estimated distance information is transmitted to the ground BS from passive UAV $k$. We assume that at time slot $t$, $I_{k,t}$ LoS or non-line-of-sight (NLoS) signal transmission paths exist between passive UAV $k$ and the ground BS.
		Then, the channel gain from passive UAV $m$ to the ground BS at time slot $t$ is
		\begin{align}\label{g_kt}
			g_{k,t}\left(\boldsymbol{q}_{k,t},{n}_{k,t}\right) & = \sum_{i = 1}^{I_{k,t}} \varepsilon_i 10^{-\frac{{L^\textrm{B}}}{10}} e^{-j \frac{2 \pi}{\lambda} \frac{n_{k,t} S \lambda}{N-1} \cos \varphi_{k,t}^i } \nonumber\\
			& = \sum_{i = 1}^{I_{k,t}} \varepsilon_i 10^{-\frac{{L^\textrm{B}}}{10}} e^{-j \frac{2 \pi S}{N-1} n_{k,t} \cos \varphi_{k,t}^i },
		\end{align}
		where $n_{k,t} \in \{1, \cdots, N\}$ is the active antenna port index of UAV $k$'s FAS at time slot $t$, $\varepsilon_i \sim {\cal {CN}} (0,1)$ is the fading coefficient of path $i$, $\varphi_{k,t}^i$ is the Angle-of-departure (AoD) of path $i$ from UAV $k$ to the BS at time slot $t$,
        and	$L^\textrm{B}$ is the distance-dependent path loss, which is given by
		\begin{equation}
			{L^\textrm{B}}=L^\textrm{F}\left(r_0\right)+10\mu\log\left(r_{k,t}\left(\boldsymbol{q}_{k,t},\boldsymbol{q}_{\textrm{B}}\right)\right)+\kappa,
		\end{equation}
		where $L^\textrm{F}\left(r_0\right)=20\log\left(r_0f_0^{\textrm B}4\pi/c\right)$ is the free-space path loss with $r_0$ being the free-space reference distance and $f_0^{\textrm B}$ being the carrier frequency, $r_{k,t}\left(\boldsymbol{q}_{k,t},\boldsymbol{q}_{\textrm{B}}\right)$ is the distance between passive UAV $k$ and the BS at time slot $t$, and $\kappa \sim {\cal N} (0, \eta^2)$ is the shadowing random variable with zero mean and $\eta^2$ variance in dB. 
		
		The signal-to-interference noise ratio (SINR) of the signal transmitted from passive UAV $k$ to the BS at time slot $t$ is
		\begin{equation}\label{SNR_BS}
			\gamma^\textrm{B}_{k,t}\left(\boldsymbol{q}_{k,t},\boldsymbol{n}_{t}\right)=\frac{p_{k,t} |g_{k,t}\left(\boldsymbol{q}_{k,t},{n}_{k,t}\right)|^2}{\sum_{k' \in \mathcal{K}\setminus\{0\}} p_{k',t} |g_{k',t}\left(\boldsymbol{q}_{k',t},{n}_{k',t}\right)|^2 + \eta^2},
		\end{equation}where $\boldsymbol{n}_t=\left[n_{1,t},\ldots,n_{4,t}\right]$ is the active antenna port vector of passive UAVs, and $p_{k,t}$ is the transmit power of passive UAV $k$ at time slot $t$. The latency of transmitting estimated distance information from the active UAV to passive UAV $k$ via the target UAV is
		\begin{equation}
			W_{k,t}^\textrm{L}\left(\boldsymbol{q}_{k,t},\boldsymbol{n}_{t}\right)=\frac{D}{B\log_2\left(1+\gamma^\textrm{B}_{k,t}\left(\boldsymbol{q}_{k,t},\boldsymbol{n}_{t}\right)\right)},
		\end{equation}where $D$ is the data size of the distance information transmitted from passive UAV $k$ to the BS, and $B$ is the bandwidth of the UAV-ground BS transmissions.
		
		\subsection{Positioning Model} 

            Let ${m}_{k,t}=d_{0,t}\left(\boldsymbol{q}_{0,t},\boldsymbol{u}_t\right)+d_{k,t}\left(\boldsymbol{u}_{t},\boldsymbol{q}_{k,t}\right)$ be the actual distance from the active UAV to passive UAV $k$ via the target UAV, and $\hat{m}_{k,t}\left(\boldsymbol{q}_{0,t},\boldsymbol{q}_{k,t}\right)$ be the distance of active-target-passive UAV links estimated by passive UAV $k$ at time slot $t$. It is assumed that an error ${e}_{k,t}\left(\gamma^\textrm{P}_{k,t}\left(\boldsymbol{q}_{0,t},\boldsymbol{q}_{k,t}\right)\right)$ exists between $\hat{m}_{k,t}\left(\boldsymbol{q}_{0,t},\boldsymbol{q}_{k,t}\right)$ and ${m}_{k,t}$ such that we have $\hat{m}_{k,t}\left(\boldsymbol{q}_{0,t},\boldsymbol{q}_{k,t}\right)={m}_{k,t}+{e}_{k,t}\left(\gamma^\textrm{P}_{k,t}\left(\boldsymbol{q}_{0,t},\boldsymbol{q}_{k,t}\right)\right)$. 
            Here, we define $\hat{\boldsymbol{m}}_t\left(\boldsymbol{\psi}_t,\boldsymbol{\theta}_t\right)=\left[\hat{m}_{1,t}\left(\boldsymbol{q}_{0,t},\boldsymbol{q}_{1,t}\right),\cdots,\hat{m}_{4,t}\left(\boldsymbol{q}_{0,t},\boldsymbol{q}_{4,t}\right)\right]$ as a vector of the distance estimated by passive UAVs, where $\boldsymbol{\psi}_{t}=\left[\psi_{0,t},\ldots,\psi_{4,t}\right]$ and $\boldsymbol{\theta}_t=\left[\theta_{0,t},\ldots,\theta_{4,t}\right]$ respectively denote the yaw angle vector and the pitch angle vector of the active UAV and passive UAVs. Each passive UAV will transmit $\hat{m}_{k,t}\left(\boldsymbol{q}_{0,t},\boldsymbol{q}_{k,t}\right)$ to the BS which will determine the location of the target UAV $\hat{\boldsymbol{u}}_{t}\left(\hat{\boldsymbol{m}}_{t}\left(\boldsymbol{\psi}_t,\boldsymbol{\theta}_t\right)\right)$ by the TDOA method \cite{TDOA}.

		\subsection{Problem Formulation}
		 Our goal is to minimize the mean squared error (MSE) between the estimated location $\hat{\boldsymbol{u}}_t\left(\hat{\boldsymbol{m}}_{t}\left(\boldsymbol{\psi}_t,\boldsymbol{\theta}_t\right)\right)$ and the ground truth location $\boldsymbol{u}_t$ of the target UAV over $T$ time slots. The minimization problem includes the trajectory optimization of controlled UAVs and antenna port selection of FAS, which can be given by
		\begin{equation}\label{optimization}
			\min_{\boldsymbol{\psi}_{t},\boldsymbol{\theta}_{t},\boldsymbol{n}_{t}} \sum_{t=1}^{T} \sqrt{\left(\hat{\boldsymbol{u}}_t\left(\hat{\boldsymbol{m}}_{t}\left(\boldsymbol{\psi}_t,\boldsymbol{\theta}_t\right)\right)-\boldsymbol{u}_{t}\right)^2},
		\end{equation}
		\begin{small}
			\begin{flalign*}
				{\rm s.t.}\quad
				&W_{k,t}^\textrm{L}\left(\boldsymbol{q}_{k,t},\boldsymbol{n}_{t}\right)\leqslant \zeta,\quad\forall{k \in \mathcal{K}\setminus\{0\}}\tag{\ref{optimization}a}\label{W_L},\\
				&\psi_{\textrm{min}}\leqslant \psi_{k,t}\leqslant \psi_{\textrm{max}}, \quad\forall{k\in\mathcal{K}}\tag{\ref{optimization}b}\label{psi},\\
				&\theta_{\textrm{min}}\leqslant \theta_{k,t}\leqslant \theta_{\textrm{max}}, \quad\forall{k\in\mathcal{K}}\tag{\ref{optimization}c}\label{theta},\\
				&n_{k,t} \in \{1, \cdots, N\}, \quad\forall{k \in \mathcal{K}\setminus\{0\}}\tag{\ref{optimization}d}\label{lkt},\\
				&{L}_{\textrm{min}}\leqslant \Vert\boldsymbol{q}_{k,t+1}-\boldsymbol{u}_{t+1}\Vert\leqslant L_{\textrm{max}}, \quad\forall{k\in\mathcal{K}} \tag{\ref{optimization}e}\label{TC},\\&L_{\textrm{min}}\leqslant \Vert\boldsymbol{q}_{k,t+1}-\boldsymbol{q}_{k',t+1}\Vert\leqslant L_{\textrm{max}}, \ \forall{k,k'\in\mathcal{K}},k\neq k' \tag{\ref{optimization}f}\label{CC},
			\end{flalign*}
		\end{small}where 
		$\psi_\textrm{min}$ and $\psi_\textrm{max}$ respectively denote the minimum and maximum yaw angle of the controlled UAVs, $\theta_\textrm{min}$ and $\theta_\textrm{max}$ represent the minimum and maximum values of the pitch angle of each controlled UAV, and $L_\textrm{min}$ and $L_\textrm{max}$ are the minimum and maximum distances between any two UAVs. \eqref{W_L} ensures that the latency of distance transmission over a UAV-ground BS link should not exceed a threshold $\zeta$. \eqref{psi} and \eqref{theta} are the constraints of the yaw angle and pitch angle that each UAV can change per time slot. \eqref{lkt} limits the number of antenna ports that each passive UAV can select. \eqref{TC} and \eqref{CC} restrict the distance between any two UAVs.

	The problem in \eqref{optimization} is challenging to solve due to the following reasons. First, the relationship among UAV trajectories ($\boldsymbol{q}_{k,t}$, $\boldsymbol{q}_{0,t}$ and $\boldsymbol{u}_t$), SNR in UAV-UAV links ($\boldsymbol{\gamma}^\textrm{P}_{k,t}$), and the estimated distance vector $\hat{\boldsymbol{m}}_t\left(\boldsymbol{\psi}_t,\boldsymbol{\theta}_t\right)$ cannot be expressed explicitly. Second, the trajectory design and antenna port selection are dependent since both of them affect the SNR of passive UAV-ground BS link. Third, the ground BS cannot obtain the real-time coordinate of the moving target UAV, while the antenna port selection and UAV trajectory design must be determined based on the target UAV trajectory. Finally, the objective function in \eqref{optimization} is non-convex since several eigenvalues of the Hessian matrix of \eqref{optimization} are negative, and the Hessian matrix of \eqref{optimization} is not positive semi-define \cite{GVK502988711}. 
		
		\section{Proposed Solution}
		In this section, we introduce an AR-MARL scheme to solve problem in \eqref{optimization}. Different from the current MARL schemes such as VD-MARL \cite{distributedhu}, the proposed AR-MARL scheme  introduces an RNN acting as a local Q function of each controlled UAV to capture its historical state-action pairs \cite{RNN_GRU_UCSD}, and a transformer that is used to analyze the importance of these historical state-action pairs \cite{vaswani2017attention}. Herein, the accuracy of global Q function approximation can be improved, thus further enhancing the positioning accuracy.
        Next, we first introduce the components of the AR-MARL algorithm. Then, the training process using the AR-MARL to optimize antenna port selections and UAV trajectories are explained.

       \begin{figure}[t!]
        \setlength\abovecaptionskip{0.cm}
        \setlength\belowcaptionskip{-0.25cm}
        \centering 
        \includegraphics[width=8.5cm]{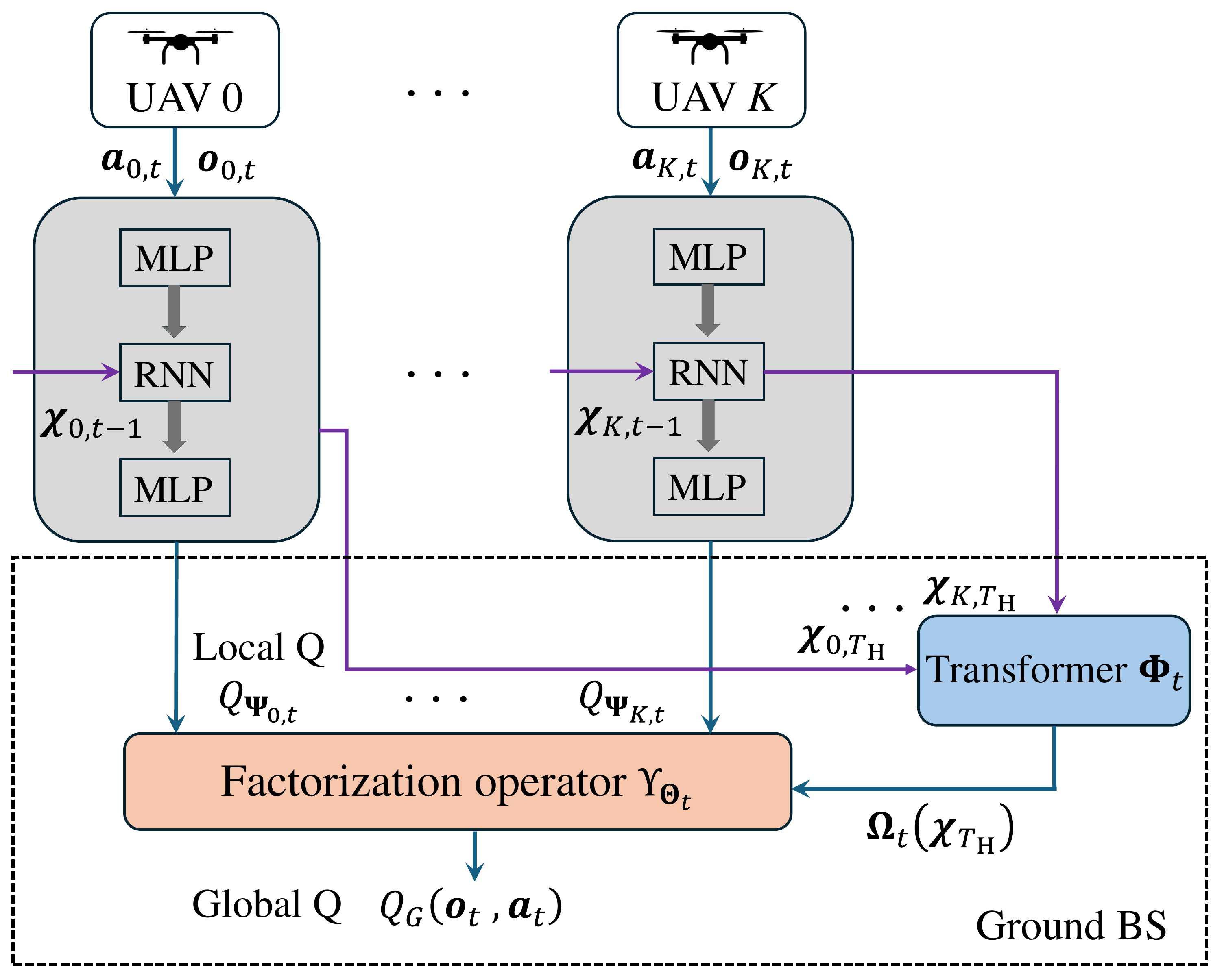}
        \captionsetup{labelsep=period} 
        \caption{Architecture of the AR-MARL scheme.}
        \label{AR-MARL} 
        \vspace{-0.4cm}
        \end{figure}
        
            \subsection{Components of the AR-MARL scheme}
		The designed AR-MARL algorithm consists of six components (see Fig. \ref{AR-MARL}): a) agents, b) states, c) actions, d) shared rewards, e) local Q function, f) attention-based coordinator, and g) global Q function, which are specified as follows:
    \begin{itemize}
        \item \emph{Agents}: The agents in the AR-MARL scheme are the active and passive UAVs. In each time slot, each active or passive UAV decides the yaw angle $\psi_{k,t}$ and pitch angle $\theta_{k,t}$ to adjust its trajectory. Meanwhile, each passive UAV $k$ selects its antenna port $n_{k,t}$ to transmit the signal to the ground BS.
        \item \emph{States}: A local state of passive UAV $k$ includes its real-time 3D location $\boldsymbol{q}_{k,t}$, the AoD vector $\boldsymbol{\varphi}_{k,t}=\left[{\varphi}^1_{k,t},\ldots,{\varphi}^i_{k,t},\ldots,{\varphi}^I_{k,t}\right]$ of $I_{n,k}$ paths from UAV $k$ to the BS at time slot $t$, and the estimated distance $\hat{m}_{k,t-1}\left(\boldsymbol{q}_{0,t-1},\boldsymbol{q}_{k,t-1}\right)$ from the active UAV to passive UAV $k$ at the previous time slot. Thus, the local state of passive UAV $k$ at time slot $t$ is expressed as $\boldsymbol{o}_{k,t}=\left[x_{k,t}, y_{k,t}, z_{k,t}, \boldsymbol{\varphi}_{k,t},\hat{m}_{k,t-1}\left(\boldsymbol{q}_{0,t-1},\boldsymbol{q}_{k,t-1}\right)\right]$. Since the active UAV does not need to estimate the distance of active-target-passive UAV links nor select the antenna port of FAS, its local state only consists of its 3D location, represented by $\boldsymbol{o}_{0,t}=\left[x_{0,t}, y_{0,t}, z_{0,t}\right]$. Hence, the joint state among all agents at time slot $t$ is a vector $\boldsymbol{o}_{t}=\left[\boldsymbol{o}_{0,t},...,\boldsymbol{o}_{4,t}\right]$.

        \item \emph{Actions}: The actions of the active UAV include the yaw angle and pitch angle adjustment that can be expressed as $\boldsymbol{a}_{0,t}=\left[\psi_{0,t},\theta_{0,t}\right]$ at time slot $t$. For each passive UAV, besides deciding the yaw angle and pitch angle, the antenna port selection of its FAS also needs to be determined. Hence, at time slot $t$, an action of passive UAV $k$ is $\boldsymbol{a}_{k,t}=\left[\psi_{k,t},\theta_{k,t},n_{k,t}\right]$. The action vector of all controlled UAVs is $\boldsymbol{a}_{t}=\left[\boldsymbol{a}_{0,t},...,\boldsymbol{a}_{4,t}\right]$.
        \item \emph{Shared Rewards}: The shared reward among the controlled UAVs is defined as the positioning accuracy of the target UAV, which is opposite to the objective function of problem \eqref{optimization}. The proposed AR-MARL will maximize this shared reward thus finding the optimal solution for the problem in \eqref{optimization}. However, if the constraints in \eqref{W_L}-\eqref{CC} are not satisfied when a UAV implements an action, a large negative penalty is assigned to discourage the undesirable action. Therefore, the shared rewards of all controlled UAVs at time slot $t$ can be given by
        \begin{equation}
            R_{t}\left(\boldsymbol{o}_t,\boldsymbol{a}_t\right)=
            \begin{cases}
            -\sqrt{\left(\hat{\boldsymbol{u}}_t\left(\hat{\boldsymbol{m}}_{t}\left(\boldsymbol{\psi}_t,\boldsymbol{\theta}_t\right)\right)-\boldsymbol{u}_{t}\right)^2}, 
            \\ \quad
            \text{if satisfying \eqref{W_L}-\eqref{CC}}, \\ 
            
            -10^6, 
            
            \text{otherwise}.
            \end{cases}
        \end{equation}
       Here, the shared reward increases as the MSE between the estimated location $\hat{\boldsymbol{u}}_t\left(\hat{\boldsymbol{m}}_{t}\left(\boldsymbol{\psi}_t,\boldsymbol{\theta}_t\right)\right)$ and the ground truth location $\boldsymbol{u}_t$ decreases, indicating that maximizing the shared reward can solve the problem in \eqref{optimization}.
       
        \item \emph{Local Q Function}: The local Q function $Q_{\boldsymbol{\Psi}_{k,t}}\left(\boldsymbol{o}_{k,t},\boldsymbol{a}_{k,t},{\boldsymbol{\chi}}_{k,t-1}\right)$ for each controlled UAV $k$ at time slot $t$ is used to estimate the cumulative shared rewards based on its local state $\boldsymbol{o}_{k,t}$, action $\boldsymbol{a}_{k,t}$, and historical state-action pair $\boldsymbol{\chi}_{k,t-1}$ at the previous time slot. In particular, the local Q function for each controlled UAV $k$ is approximated by an RNN and two multi-layer perceptrons (MLPs) with parameter vector $\boldsymbol{\Psi}_{k,t}$.
        

        \item \emph{Attention-based Coordinator}: The attention-based coordinator is used to capture the potential dependencies among historical state-action pairs of the controlled UAVs, which is realized by a transformer network. 
        This transformer network consists of multiple attention units to analyze the interactions among the controlled UAVs by weighing the importance of the input state-action pairs. In each attention unit, three MLPs are used to calculate the attention scores. Based on the calculated scores, one can determine which state-action pairs should be selected. The historical state-action pair vector $\boldsymbol{\chi}_{T_\textrm{H}}=\left<\boldsymbol{\chi}_{k,{T_\textrm{H}}}\right>_{k\in\mathcal{K}}$ of all UAVs during previous $T_\textrm{H}$ sampled time slots is the input data of the attention unit $u$. Then, the output of each attention unit is
        \begin{equation}\label{attention}
            \Omega_{u,t}(\boldsymbol{\chi}_{T_\textrm{H}}) = \text{softmax} \left( \frac{\tau_\textrm{Q} (\boldsymbol{\chi}_{T_\textrm{H}}) \tau_\textrm{K} (\boldsymbol{\chi}_{T_\textrm{H}})^\top}{\sqrt{\textrm{dim}(\Omega_{u,t}(\boldsymbol{\chi}_{T_\textrm{H}}))}} \right) \tau_\textrm{V} (\boldsymbol{\chi}_{T_\textrm{H}}),
        \end{equation}where $\tau_\textrm{Q} (\boldsymbol{\chi}_{T_\textrm{H}})$, $\tau_\textrm{K} (\boldsymbol{\chi}_{T_\textrm{H}})$, and $\tau_\textrm{V} (\boldsymbol{\chi}_{T_\textrm{H}})$ are weight matrices. $\textrm{dim}(\Omega_{u,t}(\boldsymbol{\chi}_{T_\textrm{H}}))$ is the size of output $\Omega_{u,t}(\boldsymbol{\chi}_{T_\textrm{H}})$. The output $\Omega_{u,t}(\boldsymbol{\chi}_{T_\textrm{H}})$ of all $U$ attention units will be concatenated and then passed through an MLP network. The output vector of the MLP is $\boldsymbol{\Omega}_{t}(\boldsymbol{\chi}_{T_\textrm{H}}) = \left[\Omega_{1,t}(\boldsymbol{\chi}_{T_\textrm{H}}), ..., \Omega_{u,t}(\boldsymbol{\chi}_{T_\textrm{H}}), ..., \Omega_{U,t}(\boldsymbol{\chi}_{T_\textrm{H}})\right]$.

        \item \emph{Global Q Function}: The global Q function $Q_{\textrm{G}}(\boldsymbol{o}_t, \boldsymbol{a}_t)$ is used to estimate the total rewards under a joint state $\boldsymbol{o}_t$ and an action $\boldsymbol{a}_t$ of all UAVs. Different from traditional MARL schemes that approximate a global Q function by the sum of local Q functions, we use a factorization operator $\Upsilon_{\boldsymbol{\Theta}_t}$ to aggregate local Q functions of all UAVs. The factorization operator $\Upsilon_{\boldsymbol{\Theta}_t}$ is realized with feedforward neural networks (FNNs) with parameters ${\boldsymbol{\Theta}_t}$. The local Q function $Q_{\boldsymbol{\Psi}_{k,t}}\left(\boldsymbol{o}_{k,t},\boldsymbol{a}_{k,t},{\boldsymbol{\chi}}_{k,t-1}\right)$ of each controlled UAV and the output $\boldsymbol{\Omega}_{t}(\boldsymbol{\chi}_{T_\textrm{H}})$ of the transformer network will be fed into $\Upsilon_{\boldsymbol{\Theta}_t}$ to generate the global Q function. Hence, the global Q function $Q_{\textrm{G}}(\boldsymbol{o}_t, \boldsymbol{a}_t)$ is
        \begin{align}\label{globalQ}
            Q_{\textrm{G}}(\boldsymbol{o}_t, \boldsymbol{a}_t) = \Upsilon_{\boldsymbol{\Theta}_t} \Big( & Q_{\boldsymbol{\Psi}_{1,t}}, \dots, Q_{\boldsymbol{\Psi}_{K,t}}, \boldsymbol{\Omega}_t(\boldsymbol{\chi}_{T_\textrm{H}}) \Big),
            \end{align}where $Q_{\boldsymbol{\Psi}_{k,t}}$ is short for $Q_{\boldsymbol{\Psi}_{k,t}}\left(\boldsymbol{o}_{k,t},\boldsymbol{a}_{k,t},{\boldsymbol{\chi}}_{k,t-1}\right)$, and $\boldsymbol{\Theta}_t$ is the parameter vector of ${\Upsilon}_{\boldsymbol{\Theta}_t}$.
        
       \end{itemize}

        \subsection{Training of the AR-MARL}
        Next, we introduce the training process of the AR-MARL scheme to optimize the trajectories of controlled UAVs and the antenna port selection of passive UAVs. We first explain the loss function of the AR-MARL scheme. The global loss function of the AR-MARL method is defined as the weighted MSE of temporal difference (TD) error between the global Q function $Q_{\textrm{G}}(\boldsymbol{o}_t, \boldsymbol{a}_t)$ and the target Q function $Q_{t}^{\textrm{T}}$ at time slot $t$. Different from traditional RL loss functions such as the quantile Huber loss function, the weighted MSE of TD error is flexible to adjust the importance among the historical state, action, and reward information, thus improving the positioning performance.
        The loss function of the AR-MARL is 
        \begin{equation}\label{loss_updated}
            \mathcal{L}\left(\boldsymbol{\Psi}_{0,t},..., \boldsymbol{\Psi}_{K,t},\boldsymbol{\Phi}_t,\boldsymbol{\Theta}_t\right)
            =  \sum_{t=1}^{T} w_{t} \cdot \left(Q_{\textrm{G}}(\boldsymbol{o}_t, \boldsymbol{a}_t) - Q_{t}^{\textrm{T}}\right)^2,
        \end{equation}
        where $\boldsymbol{\Phi}_t$ is a vector of the parameters of the attention-based coordinator at time slot $t$. $w_{t}$ is a weight parameter depending on the value of $Q_{\textrm{G}}(\boldsymbol{o}_t, \boldsymbol{a}_t)$, which is given by
        \begin{equation}
            w_{t} = 
                \begin{cases} 
                1, & Q_{\textrm{G}}(\boldsymbol{o}_t, \boldsymbol{a}_t) - Q_{t}^{\textrm{T}} < 0 \\
                \delta, & Q_{\textrm{G}}(\boldsymbol{o}_t, \boldsymbol{a}_t) - Q_{t}^{\textrm{T}} \geq 0
            \end{cases},
        \end{equation}where $\delta$ is a hyper-parameter to adjust the sensitivity of the model to the value of  $Q_{\textrm{G}}(\boldsymbol{o}_t, \boldsymbol{a}_t) - Q_{t}^{\textrm{T}}$.
        \newcommand{\tabincell}[2]{\begin{tabular}{@{}#1@{}}#2\end{tabular}}  

\begin{table}
\centering 
\small
 \begin{tabular}{l}
 \toprule  
 \textbf{Algorithm 1.} AR-MARL for Target UAV Positioning  \\
 \midrule  
 1: Local Q function $\boldsymbol{\Psi}_{k,t}$ of each UAV $k$, the attention-based \\ \quad  coordinator parameters $\boldsymbol{\Phi}_t$, and the parameters $\boldsymbol{\Theta}_t$ of $\Upsilon_{\boldsymbol{\Theta}_t}$.\\
 2: \textbf{for} each episode \textbf{do}\\
 3: \quad\textbf{for} each time slot $t=1,2,...,T$ \textbf{do}\\
 4: \quad\quad\textbf{for} each controlled UAV $k=1,2,...,K$ \textbf{do}\\
 5: \quad\quad\quad Observe the local state $\boldsymbol{o}_{k,t}$. \\
 6: \quad\quad\quad  Select an action $\boldsymbol{a}_{k,t}$ via $\epsilon$-greedy.\\
 7: \quad\quad\quad Calculate local Q function values.\\
 8: \quad\quad\quad Obtain the historical state-action pair $\boldsymbol{\chi}_{T_\textrm{H}}$. \\
 9: \quad\quad \textbf{end for}\\
 10:\quad\quad The BS obtains the output $\boldsymbol{\Omega}_t(\boldsymbol{\chi}_{T_\textrm{H}})$ using \eqref{attention}. \\
 11:\quad\quad The BS calculates the loss $\mathcal{L}\left(\boldsymbol{\Psi}_{0,t},..., \boldsymbol{\Psi}_{K,t},\boldsymbol{\Phi}_t,\boldsymbol{\Theta}_t\right)$\\ \quad\quad\quad based on \eqref{loss_updated} and calculates the gradients \\ \quad\quad\quad $\triangledown_{\boldsymbol{\Psi}_{k,t}}\mathcal{L}$, $\triangledown_{\boldsymbol{\Phi}_{t}}\mathcal{L}$, and $\triangledown_{\boldsymbol{\Theta}_{t}}\mathcal{L}$. \\ 
 12:\quad\quad The BS updates the attention-based coordinator \\ \quad\quad\quad parameters $\boldsymbol{\Phi}_t$ and 
  the parameters $\boldsymbol{\Theta}_t$ of $\Upsilon_{\boldsymbol{\Theta}_t}$. \\
 13:\quad\quad \textbf{for} each controlled UAV in parallel \textbf{do}\\
 14:\quad\quad\quad Update the local Q function $\boldsymbol{\Psi}_{k,t}$ using \eqref{update}. \\
 15:\quad\quad \textbf{end for} \\
 16:\quad \textbf{end for}\\
 17: \textbf{end for}\\
 18: \textbf{Output:} Parameters $\boldsymbol{\Psi}_{k,t}$, $\boldsymbol{\Phi}_t$ and $\boldsymbol{\Theta}_t$\\
\bottomrule 
\end{tabular}
\end{table}

    Given the loss function \eqref{loss_updated}, the training process is summarized as follows:
    \begin{itemize}
    \item \emph{Step 1 (Observation)}: At time slot $t$, each controlled UAV observes the considered network to determine the local state $\boldsymbol{o}_{k,t}$ and selects an action $\boldsymbol{a}_{k,t}$ based on the $\epsilon$-greedy strategy. Each controlled UAV $k$ uses historical state-action pair to calculate the local Q function $Q_{\boldsymbol{\Psi}_{k,t}}\left(\boldsymbol{o}_{k,t},\boldsymbol{a}_{k,t},{\boldsymbol{\chi}}_{k,t-1}\right)$ and predict the state-action pair $\boldsymbol{\chi}_{k,t}$. Then, the local Q function $Q_{\boldsymbol{\Psi}_{k,t}}\left(\boldsymbol{o}_{k,t},\boldsymbol{a}_{k,t},{\boldsymbol{\chi}}_{k,t-1}\right)$ and the historical state-action pair vector $\boldsymbol{\chi}_{T_\textrm{H}}$ are transmitted to the ground BS for global training in the next step. 
    \item \emph{Step 2 (Training at the BS)}: The BS approximates the global Q function $Q_{\textrm{G}}(\boldsymbol{o}_t, \boldsymbol{a}_t)$ using the factorization operator $\Upsilon_{\boldsymbol{\Theta}_t}$. The global loss $\mathcal{L}\left(\boldsymbol{\Psi}_{0,t},..., \boldsymbol{\Psi}_{K,t},\boldsymbol{\Phi}_t,\boldsymbol{\Theta}_t\right)$ is calculated based on \eqref{loss_updated} to obtain the gradients $\triangledown_{\boldsymbol{\Theta   }_{t}}\mathcal{L}\left(\boldsymbol{\Psi}_{0,t},..., \boldsymbol{\Psi}_{K,t},\boldsymbol{\Phi}_t,\boldsymbol{\Theta}_t\right)$ and $\triangledown_{\boldsymbol{\Phi}_{t}}\mathcal{L}\left(\boldsymbol{\Psi}_{0,t},..., \boldsymbol{\Psi}_{K,t},\boldsymbol{\Phi}_t,\boldsymbol{\Theta}_t\right)$. Then, the BS updates the $\boldsymbol{\Theta}_t$ of $\Upsilon_{\boldsymbol{\Theta}_t}$ in the global Q function and the transformer network $\boldsymbol{\Phi}_t$ in the attention-based coordinator.
    \item \emph{Step 3 (Training at UAVs)}: The BS periodically sends the gradient $\triangledown_{\boldsymbol{\Psi}_{k,t}}\mathcal{L}\left(\boldsymbol{\Psi}_{0,t},..., \boldsymbol{\Psi}_{K,t},\boldsymbol{\Phi}_t,\boldsymbol{\Theta}_t\right)$ from the global Q function to each controlled UAV $k$. Each controlled UAV updates its Q network $\boldsymbol{\Psi}_{k,t}$. The update of controlled UAV $k$ is
    \begin{equation}\label{update}
        \begin{split}        \boldsymbol{\Psi}_{k,t}&=\boldsymbol{\Psi}_{k,t}+\varsigma_k\triangledown_{\boldsymbol{\Psi}_{k,t}}\mathcal{L}\left(\boldsymbol{\Psi}_{0,t},..., \boldsymbol{\Psi}_{K,t},\boldsymbol{\Phi}_t,\boldsymbol{\Theta}_t\right),
        \end{split}
    \end{equation}where $\varsigma_k$ is the learning rate. The entire training process of the AR-MARL scheme is presented in Algorithm 1, where $\nabla_{\boldsymbol{\Psi}{k,t}} \mathcal{L}$ denotes the shorthand for $\nabla{\boldsymbol{\Psi}{k,t}} \mathcal{L}\left(\boldsymbol{\Psi}{0,t}, \dots, \boldsymbol{\Psi}_{K,t}, \boldsymbol{\Phi}_t, \boldsymbol{\Theta}_t\right)$. Therefore, it is obvious that the proposed AR-MARL scheme enables each controlled UAV to train its local Q function distributively.
    \end{itemize}

        \subsection{Implementation and Complexity Analysis}
        Next, we analyze the implementation and complexity of the proposed AR-MARL scheme.

        \subsubsection{Implementation Analysis} The implementation of the AR-MARL scheme for 3D UAV positioning consists of an offline training stage and an online implementation stage.
        In the offline training stage, to train the local Q function $Q_{\boldsymbol{\Psi}_{k,t}}\left(\boldsymbol{o}_{k,t},\boldsymbol{a}_{k,t},{\boldsymbol{\chi}}_{k,t-1}\right)$, the active UAV needs its real-time 3D position, and each passive UAV requires not only its 3D location but also the AoD and the estimated distance of each active-target-passive UAV link. Given this information, the active and passive UAVs select their actions such as yaw angles, pitch angles, and the antenna port selection. Based on the state and action information, as well as the global Q value, the parameters of the local Q function of each controlled UAV is updated according to \eqref{loss_updated} and \eqref{update}. Then, each passive UAV transmits the estimated distance information to the ground BS for the calculation of the global Q function $Q_{\textrm{G}}(\boldsymbol{o}_t, \boldsymbol{a}_t)$. In particular, the ground BS aggregates local Q values $Q_{\boldsymbol{\Psi}_{k,t}}\left(\boldsymbol{o}_{k,t},\boldsymbol{a}_{k,t},{\boldsymbol{\chi}}_{k,t-1}\right)$ of all UAVs using the factorization operator ${\Upsilon}_{\boldsymbol{\Theta}_t}$ in \eqref{globalQ}, and the outputs $\boldsymbol{\Omega}_{t}(\boldsymbol{\chi}_{T_\textrm{H}})$ of the transformer. In the online implementation stage, the yaw angles, pitch angles, and antenna port selections are independently determined by each controlled UAV.  

        \subsubsection{Complexity Analysis} The complexity of the AR-MARL scheme depends on the training of local Q function at each controlled UAV, and the training of the attention-based coordinator and global Q function at the BS.

        First, we analyze the complexity of training the local Q function $Q_{\boldsymbol{\Psi}_{k,t}}\left(\boldsymbol{\chi}_{k,t},\boldsymbol{a}_{k,t},{\boldsymbol{\chi}}_{k,t-1}\right)$ in each controlled UAV. Since the local Q function $Q_{\boldsymbol{\Psi}_{k,t}}\left(\boldsymbol{o}_{k,t},\boldsymbol{a}_{k,t},{\boldsymbol{\chi}}_{k,t-1}\right)$ is approximated by a gated recurrent unit (GRU) that consists of one hidden layer, the training complexity is $\mathcal{O} \left( B_\textrm{G} \left( {V_\textrm{G}}^2 + |\boldsymbol{\chi}_{k,t}| V_\textrm{G}\right)\right)$ \cite{RNNComplexity}, where $V_\textrm{G}$ is the number of neurons in the GRU hidden layer, $|\boldsymbol{\chi}_{k,t}|$ is the dimension of historical state-action pair $\boldsymbol{\chi}_{k,t}$, and $B_\textrm{G}$ is the batch size.

        Next, we analyze the complexity of training the attention-based coordinator and global Q function $Q_{\textrm{G}}(\boldsymbol{o}_t, \boldsymbol{a}_t)$ at the BS. The attention-based coordinator is approximated by a transformer network. Hence, the training complexity per iteration of the attention-based coordinator is $\mathcal{O} \left( R_\textrm{T} \left(T_\textrm{H}^2 V_\textrm{T} + T_\textrm{H} |\boldsymbol{\chi}_{T_\textrm{H}}|V_\textrm{T}+ T_\textrm{H} V_\textrm{T}^2\right) \right)$ \cite{TransformerComplexity}, where $R_\textrm{T}$ is the number of transformer network layers, $|\boldsymbol{\chi}_{T_\textrm{H}}|$ is the dimension of the transformer input $\boldsymbol{\chi}_{T_\textrm{H}}$, $V_\textrm{T}$ is is the number of neurons in each transformer hidden layer, and $T_\textrm{H}$ is the number of sampled historical time slots. 
        
        The global Q function is approximated by two FNNs in factorization operator. Hence, the training complexity of the global Q function depends on two FNNs. The complexity of the first FNN that consists one hidden layer is $\mathcal{O} \left( K V_\textrm{M}+ V_\textrm{M}^2\right)$, where $V_\textrm{M}$ is the number of neurons at the hidden layer of the first FNN, and $K$ represents the number of controlled UAVs. For the second FNN that is used to analyze the weight of the first FNN, its complexity is $\mathcal{O} \left(|\boldsymbol{\chi}_{k,t}|V_\textrm{H} + R_\textrm{H}V_\textrm{H}^2+ R_\textrm{M}V_\textrm{H}V_\textrm{M}^2 \right)$ \cite{QMIXComplexity}, where $V_\textrm{H}$ is the number of neurons of each hidden layer in the second FNN, $R_\textrm{H}$ is the the number of layers in the second FNN.
        Finally, at each iteration, the overall computation complexity of training the proposed AR-MARL scheme is $ \mathcal{O} ( B_\textrm{G} ( {V_\textrm{G}}^2 + |\boldsymbol{\chi}_{k,t}| V_\textrm{G}) + R_\textrm{T} (T_\textrm{H}^2 V_\textrm{T} + T_\textrm{H} |\boldsymbol{\chi}_{T_\textrm{H}}|V_\textrm{T}+ T_\textrm{H} V_\textrm{T}^2) K V_\textrm{M}+ V_\textrm{M}^2 + |\boldsymbol{\chi}_{k,t}|V_\textrm{H} + R_\textrm{H}V_\textrm{H}^2+ R_\textrm{M}V_\textrm{H}V_\textrm{M}^2)$.

        \section{Positioning Error Analysis and Optimization}
          In this section, we analyze how the transmit power of the active UAV and locations of the controlled UAVs affect the positioning errors of the target UAV. First, we define the relationship between the position of each controlled UAV and the estimated distance $\hat{m}_{k,t}\left(\boldsymbol{q}_{0,t},\boldsymbol{q}_{k,t}\right)$ of each active-target-passive UAV link at time slot $t$ as
          \begin{equation}\label{rela_hatm&u}
    \begin{aligned}
        \hat{m}_{k,t}\left(\boldsymbol{q}_{0,t},\boldsymbol{q}_{k,t}\right) 
        &= d_{0,t}\left(\boldsymbol{q}_{0,t},\boldsymbol{u}_t\right) 
        + d_{k,t}\left(\boldsymbol{u}_{t},\boldsymbol{q}_{k,t}\right) \\
        &\quad + {e}_{k,t}\left(\gamma^\textrm{P}_{k,t}\left(\boldsymbol{q}_{0,t},\boldsymbol{q}_{k,t}\right)\right),
    \end{aligned}
\end{equation}where the measurment error ${e}_{k,t}\left(\gamma^\textrm{P}_{k,t}\left(\boldsymbol{q}_{0,t},\boldsymbol{q}_{k,t}\right)\right)$ is a random variable that follows an independent Gaussian distribution ${\cal N}\left(0, \sigma_{k,t}^2\left(\gamma_{k,t}^\textrm{P}\right)\right)$ 
, and $\sigma_{k,t}^2\left(\gamma_{k,t}^\textrm{P}\right)$ depends on the SNR $\gamma_{k,t}^\textrm{P}$ of each active-target-passive UAV link at time slot $t$.

           Next, we calculate the differentiation of $\hat{m}_{k,t}\left(\boldsymbol{q}_{0,t},\boldsymbol{q}_{k,t}\right)$ in \eqref{rela_hatm&u}, as follows
           
          \begin{align}\label{dhatm}
          \mathrm{d}\hat{m}_{k,t} 
              & = \left(\frac{\partial d_{0,t}}{\partial x} + \frac{\partial d_{k,t}}{\partial x} + \frac{\partial e_{k,t}}{\partial x}\right) \mathrm{d}x_t \notag \\
              & + \left(\frac{\partial d_{0,t}}{\partial y} + \frac{\partial d_{k,t}}{\partial y} + \frac{\partial e_{k,t}}{\partial y}\right) \mathrm{d}y_t \notag \\
              & + \left(\frac{\partial d_{0,t}}{\partial z} + \frac{\partial d_{k,t}}{\partial z} + \frac{\partial e_{k,t}}{\partial z}\right) \mathrm{d}z_t,
          \end{align}
          where ${e}_{k,t}$ is short for ${e}_{k,t}\left(\gamma^\textrm{P}_{k,t}\left(\boldsymbol{q}_{0,t},\boldsymbol{q}_{k,t}\right)\right)$. Since $d_{0,t}=\Vert \boldsymbol{q}_{0,t}-\boldsymbol{u}_{t} \Vert$, $d_{k,t}=\Vert \boldsymbol{u}_{t} - \boldsymbol{q}_{k,t} \Vert$, we have $\frac{\partial d_{0,t}}{\partial x}=\frac{x_t - x_{k,t}}{d_{k,t}}$, and $\frac{\partial d_{k,t}}{\partial x}=\frac{x_t - x_{0,t}}{d_{0,t}}$. Thus, \eqref{dhatm} can be rewritten as

        \begin{align}\label{differentiation}
\mathrm{d}\hat{m}_{k,t} 
&= \left( \frac{x_t - x_{k,t}}{d_{k,t}}  +  \frac{x_t - x_{0,t}}{d_{0,t}} + \frac{\partial e_{k,t}}{\partial x} \right) \mathrm{d}x_t
\notag \\
&+ \left( \frac{y_t - y_{k,t}}{d_{k,t}}  +  \frac{y_t - y_{0,t}}{d_{0,t}} + \frac{\partial e_{k,t}}{\partial y} \right) \mathrm{d}y_t
\notag \\
& + \left( \frac{z_t - z_{k,t}}{d_{k,t}}  +  \frac{z_t - z_{0,t}}{d_{0,t}} + \frac{\partial e_{k,t}}{\partial z} \right) \mathrm{d}z_t, \notag \\
&\quad k = 1, 2, 3, 4,
\end{align}where $\hat{m}_{k,t}$, $d_{0,t}$, and $d_{k,t}$ are short for $\hat{m}_{k,t}\left(\boldsymbol{q}_{0,t},\boldsymbol{q}_{k,t}\right)$, $d_{0,t}\left(\boldsymbol{q}_{0,t},\boldsymbol{u}_t\right)$, and $d_{k,t}\left(\boldsymbol{u}_{t},\boldsymbol{q}_{k,t}\right)$.

        Let
        \begin{equation}\label{W}
        \boldsymbol{W} =
        \begin{bmatrix}
        \boldsymbol{X}_1 & \boldsymbol{Y}_1 & \boldsymbol{Z}_1 \\
        \boldsymbol{X}_2 & \boldsymbol{Y}_2 & \boldsymbol{Z}_2 \\
        \boldsymbol{X}_3 & \boldsymbol{Y}_3 & \boldsymbol{Z}_3 \\
        \boldsymbol{X}_4 & \boldsymbol{Y}_4 & \boldsymbol{Z}_4 \\
        \end{bmatrix},
        \end{equation}with $\boldsymbol{X}_k = \left( \frac{x_t - x_{k,t}}{d_{k,t}}  +  \frac{x_t - x_{0,t}}{d_{0,t}} + \frac{\partial e_{k,t}}{\partial x} \right)$, $\boldsymbol{Y}_k = \left( \frac{y_t - y_{k,t}}{d_{k,t}}  +  \frac{y_t - y_{0,t}}{d_{0,t}} + \frac{\partial e_{k,t}}{\partial y} \right)$, and $\boldsymbol{Z}_k = \left( \frac{z_t - z_{k,t}}{d_{k,t}}  +  \frac{z_t - z_{0,t}}{d_{0,t}} + \frac{\partial e_{k,t}}{\partial z} \right)$. Based on \eqref{W}, \eqref{differentiation} can be given by
         \begin{equation}\label{differentiation2}
             \mathrm{d}\boldsymbol{\hat{m}}_t = \boldsymbol{W} \,\mathrm{d}\boldsymbol{u}_t,
         \end{equation}   
         where $
        \mathrm{d}\boldsymbol{\hat{m}}_t = \left[\mathrm{d}\hat{m}_{1,t}, \, \mathrm{d}\hat{m}_{2,t}, \, \mathrm{d}\hat{m}_{3,t}, \, \mathrm{d}\hat{m}_{4,t}\right]^{T}$, $\mathrm{d}\boldsymbol{u}_t=\left[\mathrm{d}x_t, \mathrm{d}y_t, \mathrm{d}z_t\right]^{T}$. Given these definitions, the positioning error $\xi_t$ of the target UAV is given in the following theorem.

        \noindent \textbf{Theorem 1:} Given the 3D locations of the target UAV and all controlled UAVs, if the distances $d_{k,t}\left(\boldsymbol{u}_{t},\boldsymbol{q}_{k,t}\right)$ $\left( k = 1,2,3,4\right)$ between the target UAV and the controlled UAVs are equal, the optimized positioning error $\xi_t$ is
        
        \begin{equation}\label{xi_t}
            \xi_t = \sqrt{ \sigma_{k,t}^2 \operatorname{Tr} \left(\left(\boldsymbol{W}^T \boldsymbol{W}\right)^{-1}\right)},
        \end{equation}where $\boldsymbol{W}^T$ is the transpose of the matrix $\boldsymbol{W}$, $\left(\boldsymbol{W}^T \boldsymbol{W}\right)^{-1}$ is the inverse of the matrix $\boldsymbol{W}^T \boldsymbol{W}$, and $\operatorname{Tr}\left(\left(\boldsymbol{W}^T \boldsymbol{W}\right)^{-1}\right)$ is the trace of the matrix $\left(\boldsymbol{W}^T \boldsymbol{W}\right)^{-1}$.
        
        \noindent \emph{Proof:} Refer to Appendix A.

        From \eqref{xi_t}, we see that the positioning error of the target UAV depends on the variance $\sigma_{k,t}^\textrm{2}$ of the measurement error $e_{k,t}$, and the matrix $\boldsymbol{W}$ that depends on the 3D positions of the controlled UAVs and the target UAV. 


        Based on \textbf{Theorem 1}, we can further analyze the minimum positioning error of the target UAV for specific values of  $\sigma_{k,t}^\textrm{2}\left(\gamma_{k,t}^\textrm{P}\right)$, $\boldsymbol{u}_t$, and $\boldsymbol{q}_{0,t}$ as detailed in the following proposition.
        

        \noindent \textbf{Proposition 1:} Given the 3D locations of the target UAV $\boldsymbol{u}_t$ and the active UAV $\boldsymbol{q}_{0,t}$, and the minimum distance between the target UAV and each passive UAV, $e_{k,t}=A \times \sigma_{k,t}^2\left( \gamma_{k,t}^\textrm{P} \right)$, and $\sigma_{k,t}^\textrm{2}\left( \gamma_{k,t}^\textrm{P} \right) = \frac{1}{\gamma_{k,t}^\textrm{P}}$, the minimum positioning error of the target UAV is
        \begin{equation}
            \xi_t^\textrm{min} = \frac{3 d_{0,t} L_\textrm{min} \rho }{2 \left(\alpha_0 \beta_{k,t} \sqrt{ p_{0,t}}+ B \rho\right)}.
        \end{equation}

        \noindent \emph{Proof:} Refer to Appendix B.

        From \textbf{Proposition 1}, we see that the minimum positioning error $\xi_t^\textrm{min}$ of the target UAV depends on the minimum distance $L_\textrm{min}$ of the target-passive UAV link and the transmit power $p_{0,t}$ of the active UAV at time slot $t$. In particular, when $p_{0,t}$ increases, the positioning error decreases. 
        


        

        \begin{figure*}[t]  
            \centering
            \includegraphics[width=1.02\textwidth]{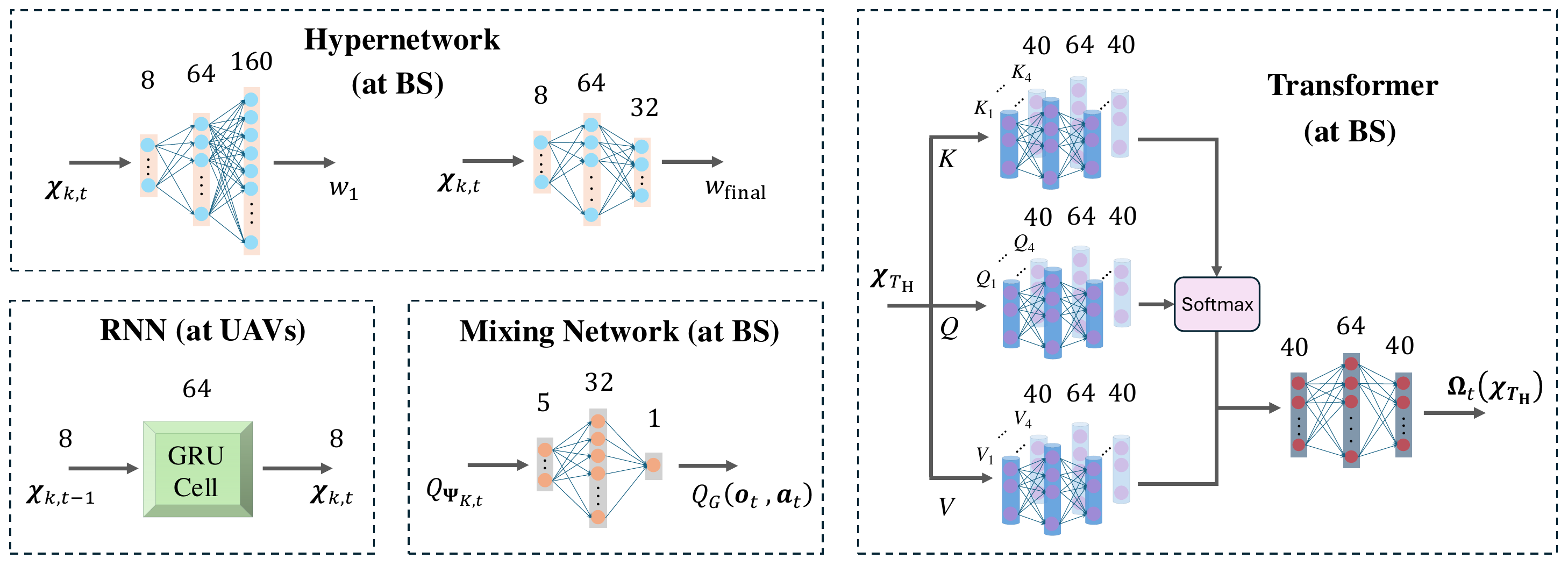}  
            \captionsetup{labelsep=period} 
            \caption{The architecture of neural networks in our proposed AR-MARL framework for simulations.}
            \label{nnstructure}  
        \end{figure*}

	\section{Simulation Results and Analysis}
	   For our simulations, the initial position of the active UAV is [300, 300, 300], and the initial positions of four FAS-assisted passive UAVs are [237, 890, 744], [310, 743, 891], [832, 497, 328], and [548, 647, 400], respectively. The position of the ground BS is [0, 0, 20]\footnote{The unit of the positions is meter.}. The flying speed $v$ of each UAV is 5 m/s. Table \ref{parameters} summarizes other parameters used in the simulations. For comparisons, we consider two baselines: a) a value decomposition based multi-agent RL (VD-MARL) algorithm \cite{distributedhu} in which each controlled UAV adjusts its trajectory and its antenna port to optimize the positioning error via decomposing the global Q function into local Q functions, and b) an independent Q learning algorithm \cite{10839261} \cite{IQL2} in which each controlled UAV utilizes a deep Q network (DQN) to optimize the trajectory and the antenna port of each controlled UAV without considering the actions of other controlled UAVs.
       
  \begin{table}[t!]
  \captionsetup{font=footnotesize} 
\caption{\label{parameters} Simulation Parameters}  
\centering 
\setlength{\abovecaptionskip}{0.cm}
\setlength{\tabcolsep}{0.8mm}{
\begin{tabular}{|c|c|c|c|}  
\hline  
$\mathbf{Parameters}$ & $\mathbf{Values}$ & $\mathbf{Parameters}$ & $\mathbf{Values}$ \\ 
\hline  
$c$ & $3\times 10^8$ m/s & $p_{k,t}$ & 5 W \\ 
\hline 
$D$ & 1000 bit & $N$ & 32 \\ 
\hline 
$S$ & 5 & $B$ & 1 MHz \\ 
\hline  
$\varphi_{k,t}^{i}$ & $[0,180^{o}]$ & $f_0^\textrm{B}$ & 2GHz  \\ 
\hline  
$L_{\textrm{min}}$ & 20 m & $L_{\textrm{max}}$ & 1000 m  \\ 
\hline  
$\psi_{\textrm{min}}$ & $-60^{o}$ & $\psi_{\textrm{max}}$ & $60^{o}$ \\ 
\hline  
$\theta_{\textrm{min}}$ & $-60^{o}$ & $\theta_{\textrm{max}}$ & $60^{o}$ \\ 
\hline 
$\rho^2$ & -90dBm & $I_{k,t}$ & $5$ \\ 
\hline 
$\zeta$ & 30ms & $T$ & $25$ \\ 
\hline 
$\varsigma_k$ & 0.05 & $B_\textrm{G}$ & $32$ \\ 
\hline 
$R_\textrm{G}$ & 1 & $K$ & $5$ \\ 
\hline 
\end{tabular}}    
\vspace{-0.1cm}
\end{table}


    \begin{figure}[t!]
    \setlength\abovecaptionskip{0.1cm}
    \setlength\belowcaptionskip{0.1cm}
    \centering 
    \includegraphics[width=8.5cm]{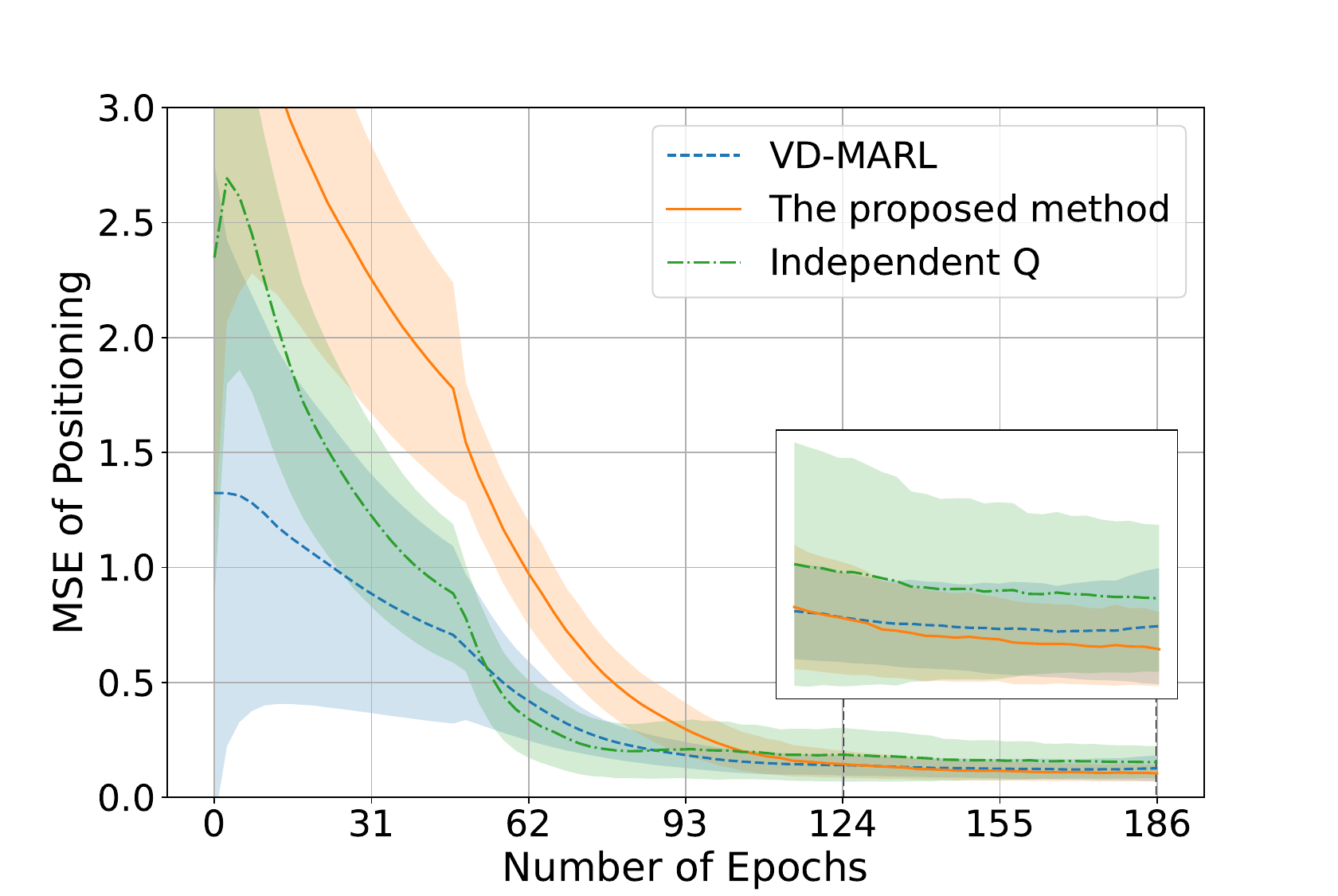}
    \captionsetup{labelsep=period} 
    \caption{MSE of positioning versus the number of training epochs (by different algorithms).}
    \label{figure_3} 
    \vspace{-0.6cm}
    \end{figure}

    \begin{figure}[t!]
    \setlength\abovecaptionskip{0.1cm}
    \setlength\belowcaptionskip{0.1cm}
    \centering 
    \includegraphics[width=8.5cm]{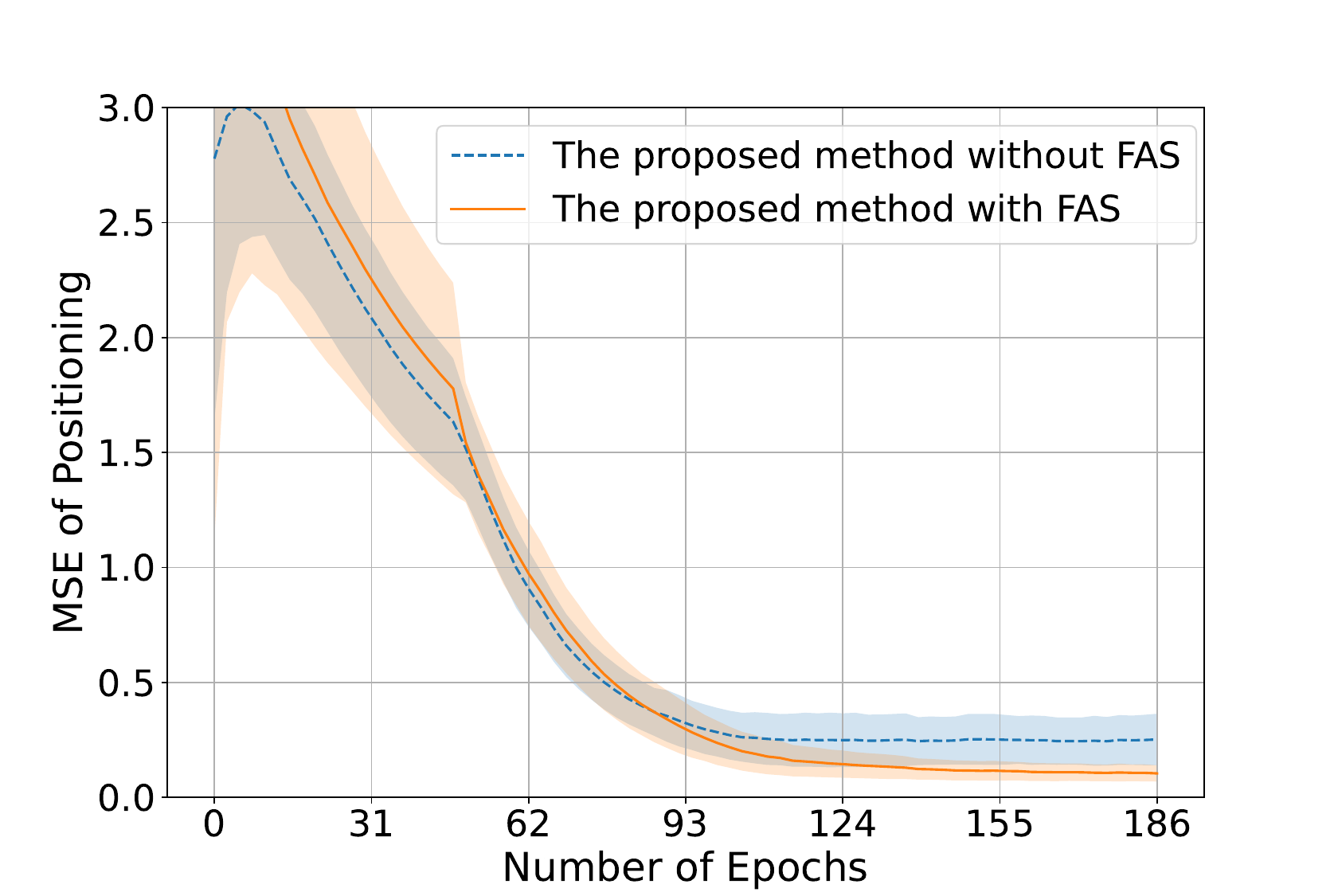}
    \captionsetup{labelsep=period} 
    \caption{MSE of positioning versus the number of training epochs (by different antenna settings).}
    \label{figure_4} 
    \vspace{-0.6cm}
    \end{figure}

    \begin{figure}[t!]
    \setlength\abovecaptionskip{0.1cm}
    \setlength\belowcaptionskip{0.1cm}
    \centering 
    \includegraphics[width=8.5cm]{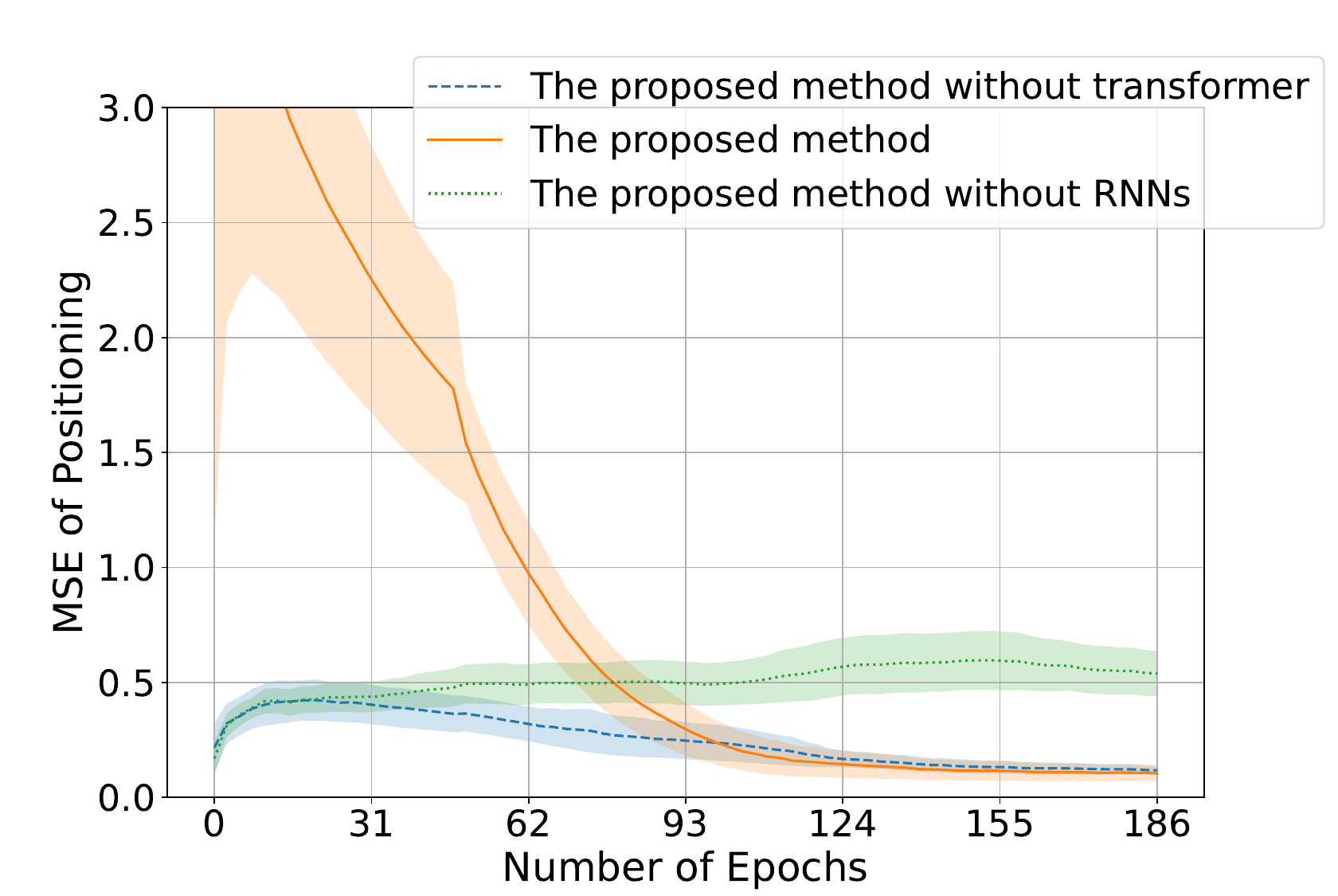}
    \captionsetup{labelsep=period} 
    \caption{MSE of positioning versus the number of training epochs (by different AR-MARL configurations).}
    \label{ARMARL_CONFIG_COMP} 
    \vspace{-0.4cm}
    \end{figure}

    

    \begin{figure}[t!]
    \setlength\abovecaptionskip{0.1cm}
    \setlength\belowcaptionskip{0.1cm}
    \centering 
    \includegraphics[width=8.5cm]{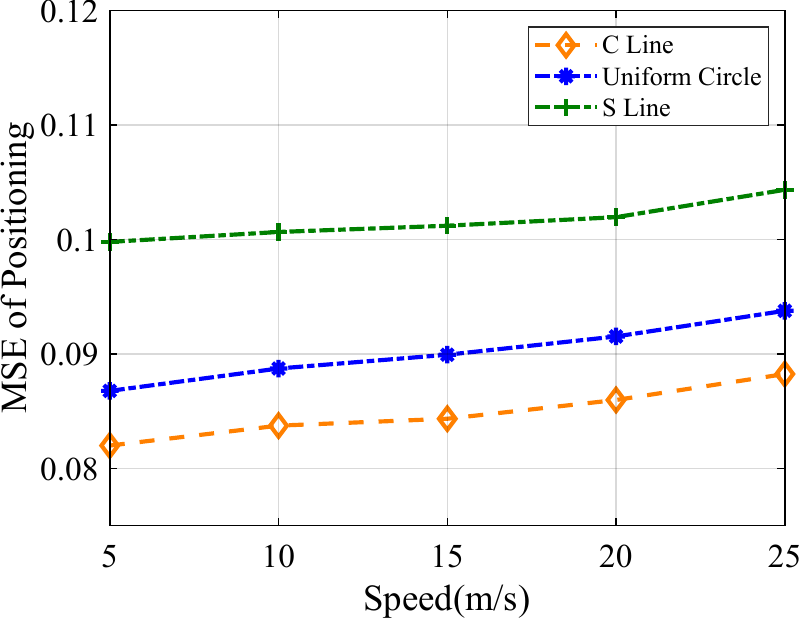}
    \captionsetup{labelsep=period} 
    \caption{MSE of positioning versus the moving speed of target UAV with different trajectories.}
    \label{speed_trajectory} 
    \vspace{-0.4cm}
    \end{figure}

    \begin{figure}[t!]
    \setlength\abovecaptionskip{0.1cm}
    \setlength\belowcaptionskip{0.1cm}
    \centering 
    \includegraphics[width=8.5cm]{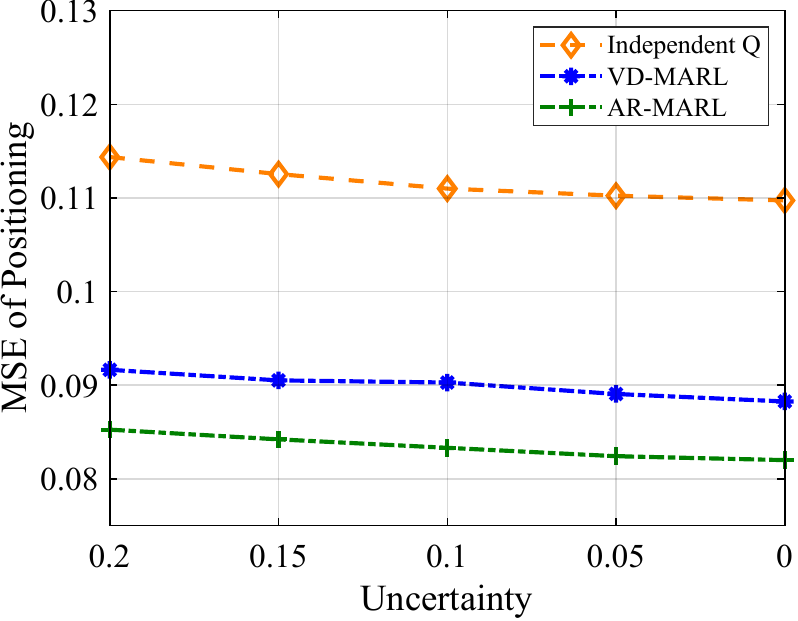}
    \captionsetup{labelsep=period} 
    \caption{MSE of positioning versus the trajectory uncertainty.}
    \label{prob} 
    \vspace{-0.4cm}
    \end{figure}

    \begin{figure}[t!]
    \setlength\abovecaptionskip{0.1cm}
    \setlength\belowcaptionskip{0.1cm}
    \centering 
    \includegraphics[width=8.5cm]{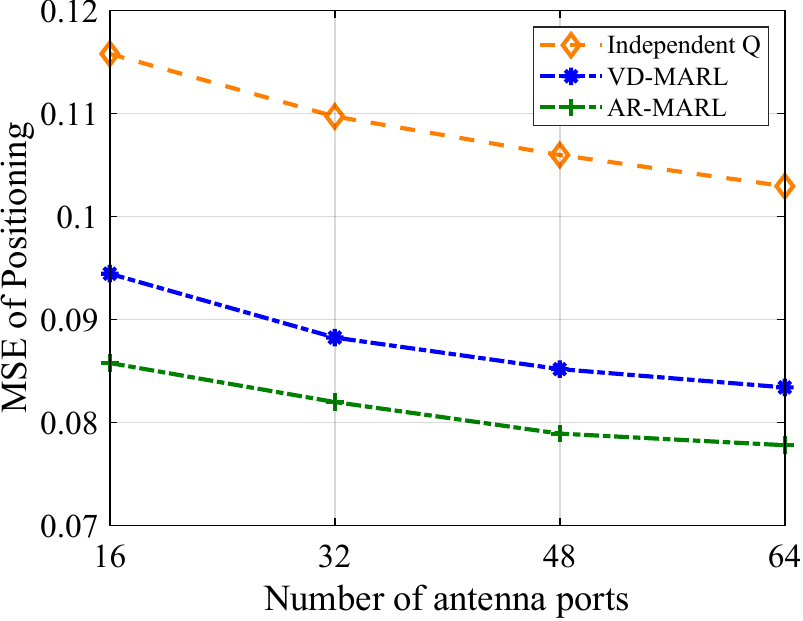}
    \captionsetup{labelsep=period} 
    \caption{MSE of positioning versus the trajectory uncertainty.}
    \label{NPort} 
    \vspace{-0.4cm}
    \end{figure}

    Fig. \ref{figure_3} shows the convergence of the considered algorithms during the training process. From Fig. \ref{figure_3}, we see that the proposed AR-MARL scheme can decrease the average positioning MSE by up to 17.5\% and 31.8\% compared to the VD-MARL and independent Q algorithm when the number of epochs is 186. The 17.5\% gain stems from the fact that proposed AR-MARL scheme uses the attention-based coordinator to capture the interdependence of the historical state-action pairs among all controlled UAVs and FNNs to nonlinearly aggregate the local Q functions for the global Q function approximation, while VD-MARL approximates the global Q function by linearly summarizing the local Q functions. The 31.8\% gain stems from the fact that our proposed AR-MARL scheme enables the UAVs to collaboratively determine their trajectories and antenna ports in order to optimize their team reward (i.e., MSE of the target UAV positioning) instead of their individual rewards.

    Fig. \ref{figure_4} shows how the antenna port selection affects the positioning performance of the proposed algorithm. In this simulation, the proposed scheme without FAS will randomly select an antenna port for each passive UAV to transmit distance information to the BS. From this figure, we see that the proposed method with FAS can reduce the average positioning MSE by up to 58.5\% compared to the proposed method without FAS when the number of epochs is 186. The reason is that using FAS enables each passive UAV to adaptively select the optimal antenna port for signal transmission, which enhances the SINR of the UAV-ground BS links.

    Fig. \ref{ARMARL_CONFIG_COMP} shows how the MSE of positioning changes as the number of epochs varies under different AR-MARL configurations. In this figure, we compare the proposed AR-MARL scheme with: i) AR-MARL scheme without using the transformer network and the global Q function is approximated only based on local Q functions; and ii) the AR-MARL scheme without using the RNNs in which each controlled UAV calculates its local Q function based on current actions and states through MLPs. From Fig. \ref{ARMARL_CONFIG_COMP}, we see that, when the number of epochs is 186, the proposed AR-MARL scheme can reduce the positioning MSE by up to 11.3\% and 80.6\% compared to the proposed scheme without transformer and the proposed scheme without RNNs, respectively. The 11.3\% gain stems from the fact that the transformer network in the attention-based coordinator can effectively analyze the importance of historical state-action pairs from each controlled UAV, and the output of the transformer will be input into the factorization factor, thus improving the approximation of the global Q function. The 80.6\% gain stems from the fact that the proposed method uses RNNs to include historical state and action information to update the local Q function, thus mitigating the short-sighted decision making and rectifies the positioning error.

    Fig. \ref{speed_trajectory} shows how the MSE of positioning varies when the speed and the movement pattern of the target UAV change. In this figure, we consider three moving trajectories of the target UAV: i) a smooth and large radius C-shaped trajectory (C Line); ii) a uniform helical trajectory (Uniform Circle); iii) a complex S-shaped trajectory (S Line). In Fig. \ref{speed_trajectory}, we see that, when the moving speed of the target UAV increases, the MSE of positioning of the target UAV with different movement patterns increases. This stems from the fact that as the target UAV moves faster, the controlled UAVs may not be able to track the target UAV accurately, thus increasing the distances between the target UAV and the controlled UAVs. Fig. \ref{speed_trajectory} also shows that the proposed AR-MARL scheme implemented under the network where target UAV moves with a C line trajectory can reduce the positioning MSE of the target UAV by up to 6.2\% compared to the algorithm implemented under the network where the target UAV moves with an uniform helical trajectory when the speed of target UAV is 15m/s. This is because the C line trajectory has less complexity compared to the other two trajectories, such that the RNN in each controlled UAV can analyze the previous state and action information to estimate the position of the target UAV. Moreover, the proposed AR-MARL scheme implemented under the network where target UAV moves with a C line trajectory can reduce the positioning MSE of the target UAV by up to 16.9\% compared to the algorithm implemented under the network where the target UAV moves with an S line trajectory when the speed of target UAV is 15m/s. This is because when the target UAV moves with the S line trajectory, it must change its yaw angle and pitch angle more frequently and rapidly. Hence, it is challenging for the controlled UAVs to estimate the position of the target UAV. 

    Fig. \ref{prob} shows how the MSE of positioning changes as the uncertainty of the target UAV trajectory varies when the target UAV moves with a C line trajectory. In this figure, the uncertainty represents the probability that the target UAV will randomly select a flying direction for left or right. For example, when the uncertainty is 0.2, target UAV has a 10\% probability of turning left by $90^{o}$, a 10\% probability of turning right by $90^{o}$, and 80\% probability of moving with the C line trajectory. From Fig. \ref{prob}, we see that as the uncertainty in the target UAV trajectory decreases, the MSE of positioning of the target UAV decreases among all the schemes. This is because as the uncertainty of the trajectory decreases, the controlled UAVs can observe the states more accurately, which simplifies the decision process and helps to better decide their trajectories and antenna port selections. Fig. \ref{prob} also shows that the proposed scheme can reduce the MSE of positioning by up to 7.8\% compared to the VD-MARL scheme when the uncertainty is 0.1. This is due to the fact that the proposed AR-MARL scheme can use RNN in each local Q function and the attention-based coordinator to analyze the historical and global state and action information, which helps each controlled UAV to learn the uncertainty of the environment and make better decisions.
    Fig.~\ref{prob} shows that the proposed scheme can reduce the MSE of positioning by up to 25.6\% compared to the independent Q scheme when the uncertainty is 0.2. This gain stems from the fact that our proposed scheme allows controlled UAVs to collaboratively determine their trajectories and antenna ports in order to optimize their shared rewards, and share the local Q function of each controlled UAV to calculate the global Q function, which enhance the cooperation among the controlled UAVs.

    Fig. \ref{NPort} shows how the MSE of the target UAV positioning changes as the number of antenna ports varies when the target UAV moves with a C line trajectory. From this figure, we see that as the number of antenna ports at each passive UAV increases, the positioning MSE resulting from all considered schemes decreases. This is because as the number of antenna ports increases, the passive UAVs can find a better antenna port to improve the SINR of the UAV-ground BS links for signal transmission. Fig. \ref{NPort} also shows that the proposed AR-MARL scheme can reduce the MSE of positioning by up to 9.2\% compared to the VD-MARL scheme when the number of antenna ports is 16. This gain stems from the fact that the proposed AR-MARL scheme uses RNNs in each local Q function and the attention-based coordinator to analyze historical antenna port selections and the observed AoD information.

		\section{Conclusion}
		In this paper, we have proposed a novel FAS-assisted 3D passive UAVs positioning framework that uses an active UAV and four FAS-assisted passive UAVs to cooperatively localize a target UAV. This positioning problem has been formulated as an optimization problem, whose goal is to minimize the average positioning error of the target UAV by optimizing the trajectories of all controlled UAVs and the antenna ports of passive UAVs. To address this problem, we have proposed an AR-MARL scheme that uses an RNN in each controlled UAV to capture its historical state-action pairs, and a transformer network at the BS to analyze the importance of these historical state-action pairs. Through this combination of RNN and transformer networks, the proposed scheme can achieve a more accurate approximation of the global Q value, ultimately enhancing the positioning accuracy of the target UAV.
        Simulation results have shown that the proposed AR-MARL scheme significantly outperforms the baselines with regards to the average positioning accuracy. 

        \appendix 
        
        \setcounter{section}{0}
        
        \renewcommand{\thesection}{A}
        \subsection{Proof of the Theorem 1}
        According to \eqref{differentiation2}, the positioning error $\xi_t$ between the estimated location $\hat{\boldsymbol{u}}_t\left(\hat{\boldsymbol{m}}_{t}\left(\boldsymbol{\psi}_t,\boldsymbol{\theta}_t\right)\right)$ and the ground truth location $\boldsymbol{u}_t$ of the target UAV can be expressed by $\xi_t=\sqrt{(\mathrm{d}x_t)^2 + (\mathrm{d}y_t)^2 + (\mathrm{d}z_t)^2}=\sqrt{\operatorname{Tr}\left(\mathbb{E}\left[\mathrm{d}\boldsymbol{u}_t \mathrm{d}\boldsymbol{u}_t^{T}\right]\right)}$ \cite{5466526} with $\operatorname{Tr}(\cdot)$ being the trace of the matrix $\mathbb{E}\left[\mathrm{d}\boldsymbol{u}_t \mathrm{d}\boldsymbol{u}_t^{T}\right]$ and $(\cdot)^T$ being the transpose of the matrix $\mathrm{d}\boldsymbol{u}_t$.
        Next, we calculate the value of $\mathbb{E}\left[\mathrm{d}\boldsymbol{u}_t \mathrm{d}\boldsymbol{u}_t^{T}\right]$, we first rewrite \eqref{differentiation2} as
        \begin{equation}\label{inversedifferentiation}
            \mathrm{d}\boldsymbol{u}_t = \left( \boldsymbol{W}^T \boldsymbol{W} \right)^{-1} \boldsymbol{W}^T \mathrm{d}{\hat{\boldsymbol{m}}}_t.
        \end{equation}
        
        Then, the value of $\mathbb{E}\left[\mathrm{d}\boldsymbol{u}_t \mathrm{d}\boldsymbol{u}_t^{T}\right]$ can be calculated by

        $\mathbb{E} \left[\mathrm{d}\boldsymbol{u}_t \mathrm{d}\boldsymbol{u}_t^T\right]$
        \begin{align} \label{EU} 
&= \mathbb{E} \left[
\left(\boldsymbol{W}^T \boldsymbol{W}\right)^{-1} \boldsymbol{W}^T \mathrm{d}\hat{\boldsymbol{m}}_t 
\left(\left(\boldsymbol{X}^T \boldsymbol{W}\right)^{-1} \boldsymbol{W}^T \mathrm{d}\hat{\boldsymbol{m}}_t\right)^T
\right] \notag \\
&= \mathbb{E} \left[
\left(\boldsymbol{W}^T \boldsymbol{W}\right)^{-1} \boldsymbol{W}^T \mathrm{d}\hat{\boldsymbol{m}}_t \mathrm{d}\hat{\boldsymbol{m}}_t^T 
\left(\left(\boldsymbol{W}^T \boldsymbol{W}\right)^{-1} \boldsymbol{W}^T\right)^T
\right] \notag \\
&= \left(\boldsymbol{W}^T \boldsymbol{W}\right)^{-1} \boldsymbol{W}^T 
\mathbb{E}\left[\mathrm{d}\hat{\boldsymbol{m}}_t \mathrm{d}\hat{\boldsymbol{m}}_t^T\right] 
\left(\left(\boldsymbol{W}^T \boldsymbol{W}\right)^{-1} \boldsymbol{W}^T\right)^T.
\end{align}
        Since $\mathbb{E}\left[\mathrm{d}\hat{\boldsymbol{m}}_t \mathrm{d}\hat{\boldsymbol{m}}_t^T\right]$ is the covariance matrix of vector $\mathrm{d}\hat{\boldsymbol{m}}_t$ with zero mean, and $\mathrm{d}\hat{m}_{k,t}$ is independent with each other, we have $\mathbb{E}\left[\mathrm{d}\hat{\boldsymbol{m}}_t \mathrm{d}\hat{\boldsymbol{m}}_t^T\right]=\textrm{diag}\left(\sigma_{1,t}^\textrm{2}, \sigma_{2,t}^\textrm{2}, \sigma_{3,t}^\textrm{2}, \sigma_{4,t}^\textrm{2}\right)$, where $\sigma_{k,t}^\textrm{2}$ is the variance of independent Gaussian measurement error $e_{k,t}$ of each passive UAV at time slot $t$ \cite{1981TASSP}. 
        
        If the distance between the target UAV and each controlled UAV is equal, we have $\sigma_{1,t}^\textrm{2} = \sigma_{2,t}^\textrm{2} = \sigma_{3,t}^\textrm{2} = \sigma_{4,t}^\textrm{2}$.
        Hence, $\mathbb{E}\left[\mathrm{d}\hat{\boldsymbol{m}}_t \mathrm{d}\hat{\boldsymbol{m}}_t^T\right]$ in \eqref{EU} can be rewritten as  
        \begin{equation} \label{EM}       \mathbb{E}\left[\mathrm{d}\hat{\boldsymbol{m}}_t \mathrm{d}\hat{\boldsymbol{m}}_t^T\right] = \sigma_{k,t}^\textrm{2}\boldsymbol{D},
        \end{equation}
        where $\boldsymbol{D}=\textrm{diag}\left(1,1,1,1\right)$. Then, we substitute \eqref{EM} into \eqref{EU} as
        \begin{align}\label{E3}
        \mathbb{E} \left[\mathrm{d}\boldsymbol{u}_t \mathrm{d}\boldsymbol{u}_t^T\right] &= \sigma_{k,t}^\textrm{2} \left(\boldsymbol{W}^T \boldsymbol{W}\right)^{-1} \boldsymbol{W}^T \left(\left(\boldsymbol{W}^T \boldsymbol{W}\right)^{-1} \boldsymbol{W}^T\right)^T \notag \\
    &= \sigma_{k,t}^\textrm{2} \left(\boldsymbol{W}^T \boldsymbol{W}\right)^{-1} \boldsymbol{W}^T \boldsymbol{W} \left(\boldsymbol{W}^T \boldsymbol{W}\right)^{-1} \notag \\
    &= \sigma_{k,t}^\textrm{2}\left(\boldsymbol{W}^T \boldsymbol{W}\right)^{-1}.
\end{align}

        
        According to \eqref{E3}, the positioning error $\xi_t$ of the target UAV can be expressed as

        \begin{align}\label{E4}
        \xi_t &=
             \sqrt{\operatorname{Tr} \left( \sigma_{k,t}^\textrm{2} \left(\boldsymbol{W}^T \boldsymbol{W}\right)^{-1} \right)} = \sqrt{ \sigma_{k,t}^\textrm{2} \operatorname{Tr} \left(\left(\boldsymbol{W}^T \boldsymbol{W}\right)^{-1}\right)}. 
        \end{align}

        This completes the proof.

        
        
        \setcounter{section}{0}
        \renewcommand{\thesection}{B}
        \subsection{Proof of the Proposition 1}      
        As $\sigma_{k,t}^\textrm{2} = \frac{1}{\gamma_{k,t}^\textrm{P}} = a \left(d_{0,t}\left(\boldsymbol{q}_{0,t},\boldsymbol{u}_t\right) d_{k,t}\left(\boldsymbol{u}_{t},\boldsymbol{q}_{k,t}\right) \right)^2$, the positioning error $\xi_t$ can be rewritten as

        \begin{equation}\label{error2}
            \xi_t = \sqrt{ a \left(d_{0,t}\left(\boldsymbol{q}_{0,t},\boldsymbol{u}_t\right) d_{k,t}\left(\boldsymbol{u}_{t},\boldsymbol{q}_{k,t}\right) \right)^2 \operatorname{Tr} \left(\left(\boldsymbol{W}^T \boldsymbol{W}\right)^{-1}\right)},
        \end{equation}where $a = \frac{\rho^2}{p_{0,t} \beta_{k,t}^2 \alpha_0^2}$.
        
        Next, we calculate $\operatorname{Tr} \left(\boldsymbol{W}^T \boldsymbol{W}\right)$, we first calculate the value of $\boldsymbol{W}$ that depends on $e_{k,t}$. In particular, since $e_{k,t} = B \sigma_{k,t}$, and $\sigma_{k,t} = \sqrt{\frac{1}{\gamma_{k,t}^\textrm{P}}}$, we have $e_{k,t} = B \sqrt{\frac{1}{\gamma_{k,t}^\textrm{P}}}  = B\frac{\rho d_{0,t} d_{k,t}}{\alpha_0 \beta_{k,t} \sqrt{p_{0,t}}}$.
        
        Then, the estimated distance $\hat{m}_{k,t}\left(\boldsymbol{q}_{0,t}, \boldsymbol{q}_{k,t}\right)$ in \eqref{rela_hatm&u} is

        \begin{equation}\label{hatm}
           \begin{aligned}
\hat{m}_{k,t}\left(\boldsymbol{q}_{0,t},\boldsymbol{q}_{k,t}\right) &= d_{0,t}\left(\boldsymbol{q}_{0,t},\boldsymbol{u}_t\right) + d_{k,t}\left(\boldsymbol{u}_{t},\boldsymbol{q}_{k,t}\right)  \\
           &+ B\frac{\rho d_{0,t}\left(\boldsymbol{q}_{0,t},\boldsymbol{u}_t\right)  d_{k,t}\left(\boldsymbol{u}_{t},\boldsymbol{q}_{k,t}\right)}{\alpha_0 \beta_{k,t} \sqrt{p_{0,t}}}.
           \end{aligned}
           \end{equation}
           
        Given the 3D locations of the target UAV (i.e., $\boldsymbol{u}_t$) and active UAV (i.e., $\boldsymbol{q}_{0,t}$) respectively, the distance $d_{0,t}\left(\boldsymbol{q}_{0,t},\boldsymbol{u}_t\right)$ of the active-target UAV link is a constant. Then, the derivative of \eqref{hatm} is
        \begin{align}\label{dm}
\mathrm{d}\hat{m}_{k,t} 
&= \left( \frac{x_t - x_{k,t}}{d_{k,t}}  + \frac{x_t - x_{k,t}}{d_{k,t}} \cdot \frac{B \rho }{\alpha_0 \beta_{k,t} \sqrt{p_{0,t}}} \right) \mathrm{d}x_t
\notag \\
&+ \left( \frac{y_t - y_{k,t}}{d_{k,t}}  + \frac{y_t - y_{k,t}}{d_{k,t}} \cdot \frac{B \rho }{\alpha_0 \beta_{k,t} \sqrt{p_{0,t}}} \right) \mathrm{d}y_t
\notag \\
& + \left( \frac{z_t - z_{k,t}}{d_{k,t}}  + \frac{z_t - z_{k,t}}{d_{k,t}} \cdot \frac{B \rho }{\alpha_0 \beta_{k,t} \sqrt{p_{0,t}}} \right) \mathrm{d}z_t.
\end{align}

        Based on \eqref{dm}, the matrix $\boldsymbol{W}$ can be rewritten as
        \begin{equation}\label{W_updated}
        \boldsymbol{W} = \frac{1 + \frac{B \rho }{\alpha_0 \beta_{k,t} \sqrt{p_{0,t}}}}{d_{k,t}}
        \begin{bmatrix}
        x_1 - x_{1,t} & y_1 - y_{1,t} & z_1 - z_{1,t} \\
        x_2 - x_{2,t} & y_2 - y_{2,t} & z_2 - z_{2,t} \\
        x_3 - x_{3,t} & y_3 - y_{3,t} & z_3 - z_{3,t} \\
        x_4 - x_{4,t} & y_4 - y_{4,t} & z_4 - z_{4,t} \\
        \end{bmatrix}.
        \end{equation}

        Hence, $\operatorname{Tr} \left(\boldsymbol{W}^T \boldsymbol{W}\right)$ is calculated as

        \begin{align}\label{Tr_Cal}        \operatorname{Tr}\left(\boldsymbol{W}^T \boldsymbol{W}\right) &= \frac{\left(1+\frac{B \rho }{\alpha_0 \beta_{k,t} \sqrt{p_{0,t}}}\right)^2}{d_{k,t}^2} \left( \sum_{k=1}^{4} (x_t - x_{k,t})^2 \right. \notag \\ 
        &\quad+ \left. \sum_{k=1}^{4} (y_t - y_{k,t})^2 + \sum_{k=1}^{4} (z_t - z_{k,t})^2 \right) \notag \\
        &= \frac{\left(1+\frac{B \rho }{\alpha_0 \beta_{k,t} \sqrt{p_{0,t}}}\right)^2}{d_{k,t}^2} \left( \sum_{k=1}^{4} d_{k,t}^2 \right) \notag \\
        &= 4 \left(1+\frac{B \rho }{\alpha_0 \beta_{k,t} \sqrt{p_{0,t}}}\right)^2.
\end{align}
        
        From \eqref{error2} and \eqref{Tr_Cal}, we see that the positioning error depends on $\operatorname{Tr} \left(\left(\boldsymbol{W}^T \boldsymbol{W}\right)^{-1}\right)$. However, we only know the value of $\operatorname{Tr} \left(\boldsymbol{W}^T \boldsymbol{W}\right)$. Therefore, we need to use $\operatorname{Tr} \left(\boldsymbol{W}^T \boldsymbol{W}\right)$ in \eqref{Tr_Cal} to present $\operatorname{Tr} \left(\left(\boldsymbol{W}^T \boldsymbol{W}\right)^{-1}\right)$ in \eqref{error2}. Since $\operatorname{Tr} \left(\left(\boldsymbol{W}^T \boldsymbol{W}\right)^{-1}\right) = 
        \sum_{w=1}^{3} \frac{1}{\varsigma_w}$ with $\varsigma_w$ being the eigenvalue of $\boldsymbol{W}^T \boldsymbol{W}^{-1}$ \cite{2009_JSTSP}, the positioning error $\xi_t$ is rewritten as
        \begin{align}\label{layout_error}
            \xi_t &= \sqrt{a \left(  d_{0,t} d_{k,t}\right)^2 \sum_{w=1}^{3} \frac{1}{\varsigma_w}} \notag \\
&\overset{\textrm{(a)}}{\geq} \sqrt{a \left( d_{0,t} d_{k,t}\right)^2 3 \left(\prod_{w=1}^{3} \frac{1}{\varsigma_w}\right)^{\frac{1}{3}}} \notag \\
&\overset{\textrm{(b)}}= \sqrt{a \left( d_{0,t} d_{k,t}\right)^2 3 \left(\frac{3}{\operatorname{Tr}\left(\boldsymbol{W}^T \boldsymbol{W}\right)}\right)}, 
        \end{align}where the inequality equation $\left(\textrm{a}\right)$ is hold only when $\varsigma_1 = \varsigma_2 = \varsigma_3$ according to the triangle-inequality. $\varsigma_1 = \varsigma_2 = \varsigma_3$ is satisfied in this proof since we already assumed that the distance between the target UAV and each passive UAV are the same. Equation $\left(\textrm{b}\right)$ derives from the fact that $\varsigma_1 + \varsigma_2 + \varsigma_3 = \operatorname{Tr}\left(\boldsymbol{W}^T \boldsymbol{W}\right)$ and $\varsigma_w = \frac{1}{3}\operatorname{Tr}\left(\boldsymbol{W}^T \boldsymbol{W}\right)$ when $\varsigma_1 = \varsigma_2 = \varsigma_3$.

        Finally, substituting \eqref{Tr_Cal} into \eqref{layout_error}, the positioning error can be given by
        \begin{align}
            \xi_t & =  \sqrt{3a\left(d_{0,t}d_{k,t}\right)^2 \left(\frac{3}{4 \left(1+\frac{B \rho }{\alpha_0 \beta_{k,t} \sqrt{p_{0,t}}}\right)^2}\right)} \notag \\
        &= \frac{3 d_{0,t} d_{k,t} \rho }{2 \left(\alpha_0 \beta_{k,t} \sqrt{ p_{0,t}}+ B \rho\right)}  
        \overset{\textrm{(c)}}{\geq} \frac{3 d_{0,t} L_\textrm{min} \rho }{2 \left(\alpha_0 \beta_{k,t} \sqrt{ p_{0,t}}+ B \rho\right)},
        \end{align}
        where equation $\left(\textrm{c}\right)$ derives from the fact that $d_{k,t} \geq L_\textrm{min}$ as represented in \eqref{TC}.
        This completes the proof.
        

		\bibliographystyle{IEEEtran}
		\bibliography{IEEEabrv,ref}
		
	\end{CJK}
\end{document}